RESEARCH ARTICLE

# The soil seed bank can buffer long-term compositional changes in annual plant communities

Niv DeMalach[1,2*], Jaime Kigel[2], Marcelo Sternberg[1]

[1] School of Plant Sciences and Food Security, George S. Wise Faculty of Life Sciences, Tel Aviv University, Tel Aviv, Israel

[2] Institute of Plant Sciences and Genetics in Agriculture, Robert H. Smith Faculty of Agriculture, Food and Environment, Hebrew University of Jerusalem, Rehovot, Israel

* Corresponding author: Nivdemalach@mail.huji.ac.il

**ABSTRACT**

1. Ecological theory predicts that the soil seed bank stabilises the composition of annual plant communities in the face of environmental variability. However, long-term data on the community dynamics in the seed bank and the standing vegetation are needed to test this prediction.
2. We tested the hypothesis that the composition of the seed bank undergoes lower temporal variability than the standing vegetation in a nine-year study in Mediterranean, semi-arid, and arid ecosystems. The composition of the seed bank was estimated by collecting soil cores from the studied sites on an annual basis. Seedling emergence under optimal watering conditions was measured in each soil core for three consecutive years, to account for seed dormancy.
3. In all sites, the composition of the seed bank differed from the vegetation throughout the years. Small-seeded and dormant-seeded species had a higher frequency in the seed bank than in the standing vegetation. In contrast, functional group membership (grasses vs. forbs) did not explain differences in species frequency between the seed bank and the vegetation after controlling for differences between grasses and forbs in seed mass and seed dormancy.
4. Contrary to predictions, the magnitude of year-to-year variability (the mean compositional dissimilarity between consecutive years) was not lower in the seed bank than in the vegetation in all sites. However, long-term compositional trends in the seed bank were weaker than in the vegetation in the Mediterranean and semi-arid sites. In the arid site where year-to-year variability was highest, no long-term trends were observed.

5. Synthesis: The effect of the seed bank on the temporal variability of the vegetation in annual communities depends on site conditions and time scale. While the year-to-year variability of the seed bank is similar to the vegetation, the soil seed bank can buffer long-term trends.

## INTRODUCTION

Understanding the factors driving community stability is a key goal in ecology (Cleland et al., 2013; Collins, 2000; de Mazancourt et al., 2013; Komatsu et al., 2019). This goal is increasingly important in times of abrupt shifts in species composition driven by climate and land-use changes (Harrison, Gornish, & Copeland, 2015; Liu et al., 2018; Song et al., 2018; Swenson, Hulshof, Katabuchi, & Enquist, 2020). For plant communities, the natural storage of seeds in the soil (hereafter seed bank) is considered essential for compositional stability because seeds are highly resistant to environmental hazards (Angert, Huxman, Chesson, & Venable, 2009; Cohen, 1966; Ooi, 2012).

Soil seed banks are especially important in ecosystems with high rainfall variability, such as drylands (Huang, Yu, Guan, Wang, & Guo, 2016; Kigel, 1995). Currently, drylands cover 45% of the world's land surface (Prăvălie, 2016) and their cover is predicted to increase to 56% by the end of this century (Huang et al., 2016). Many drylands are dominated by annual plants that germinate each year from the seed bank (Angert et al., 2009; Tielborger et al., 2014). These communities are frequently characterized by high temporal variability in species composition driven by asynchronized fluctuations among populations of coexisting species (Bar-Massada & Hadar, 2017; Hobbs, Yates, & Mooney, 2007). Theoretically, the seed bank can buffer two types of compositional changes, namely year-to-year variability (Cohen, 1966) and long-term community changes (Koopmann, Müller, Tellier, & Živković, 2017). Year-to-year variability may result from unpredictable differences among years in environmental conditions (e.g. precipitation, temperature). Long-term community changes are often caused by a trended variation in environmental conditions or management practices. Climate change models predict changes in both the mean and the variance of climatic conditions which will probably affect both year-to-year variability and long-term trends (Donat, Lowry, Alexander, O'Gorman, & Maher, 2016; Huang et al., 2016). Similarly, global land-use changes lead to directional changes in community composition (e.g. succession) but also affect year-to-year variability (Allan et al., 2014).

The role of the soil seed bank in buffering year-to-year environmental variability can vary across ecosystems. Classical theory predicts that a higher dormancy fraction will be favored in systems with high rainfall uncertainty such as deserts, while lower dormancy will be favored in more predictable environments (Cohen, 1966; Venable & Brown, 1988). However, the persistence of seeds in the soil is affected not only by dormancy but also by other factors such as seed predation, pathogen attack, and mechanical decay (Kigel, 1995; Thompson, 1987).

While ecological theory highlights the role of the seed bank in stabilising plant communities (Cohen, 1966; Venable & Brown, 1988), long-term monitoring of seed bank dynamics are scarce. Most empirical studies have focused on the short-term dynamics (<3 years) of seed banks (Bossuyt & Honnay, 2008; Osem, Perevolotsky, & Kigel, 2006) while several studies have used chronosequences as a substitute for the lack of long-term data from the same location (Dalling & Denslow, 1998; Török et al., 2018). We know of only one study that analyzed long-term seed bank dynamics, focusing on the ten most abundant species within a desert annual community (Venable & Kimball, 2012). Here, we compared temporal compositional trends in the seed bank and ensuing standing vegetation in annual plant communities spanning Mediterranean, semi-arid and arid ecosystems.

We hypothesized that year-to-year variability in the composition of the vegetation will increase with increasing aridity (because rainfall variability increases with aridity) while the seed bank will be more stable (Cohen, 1966; Venable & Brown, 1988), i.e. the role of the seed bank in buffering year-to-year variability will increase with aridity. Additionally, assuming that the seed bank is a major driver of the high stability of Middle-Eastern communities (Sternberg et al., 2015; Tielborger et al., 2014), we predicted that the seed bank will experience weaker long-term compositional trends than the vegetation.

A further aim of the study was to explain the differences in composition between the seed bank and the vegetation using a trait-based approach. Small-seeded species typically have higher fecundity ('the size-number tradeoff', Jakobsson & Eriksson, 2000), and higher persistence in the soil (Funes, Basconcelo, Díaz, & Cabido, 1999; Thompson, Band, & Hodgson, 1993; Thompson, Bakker, Bekker, & Hodgson, 1998). However, small-seeded species often have lower survival at the seedling stage (Ben-Hur, Fragman-Sapir, Hadas, Singer, & Kadmon, 2012; Metz et al., 2010). Therefore, we predicted that small-seeded species will be relatively more common in the seed bank than in the vegetation. We also hypothesized that species with higher seed dormancy will be more common in the seed bank (Thompson, 1987)

## METHODS

### Study sites

The study was conducted at three sites located along a rainfall gradient (ca. 100 km length) in Israel. All sites were located over the same calcareous bedrock on south-facing slopes at similar altitudes and experienced similar mean annual temperatures that range from 17.7 to 19.1 °C. The length of the growing season is determined by the rainfall, usually commencing in October–November and ending in April–May, with shorter seasons in drier sites. A detailed description of the sites appears in previous publications (Harel, Holzapfel, & Sternberg, 2011; Tielborger et al., 2014).

Briefly, the three sites represent three different climatic regions: Mediterranean (Matta LTER; N 31$^{o}$ 42'; E 35$^{o}$ 03'), semi-arid (N 31$^{o}$23'; E 34$^{o}$54'), and arid (N 30$^{o}$ 52', E 34$^{o}$ 46'). Thus, the sites have relatively low species overlap in terms of Jaccard's similarity (Mediterranean–semi-arid: 0.64, Mediterranean–arid: 0.18, and semi-arid–arid: 0.22, see Tables S1-S3 for full species lists). The long-term mean annual rainfall in these three sites is 540, 300, and 90 mm with a coefficient of variation (CV) of 30%, 37%, and 51% respectively (Tielborger et al., 2014). The mean annual rainfall during the years of the study (2000/2001–2009/2010) was 502, 245, and 79 mm with a CV of 24%, 32%, and 48% respectively. All sites were fenced against grazing (by sheep and goats) in 2001. Before the establishment of the experimental plots, grazing intensity was high in the semi-arid site, intermediate in the Mediterranean site, and negligible in the arid site (M. Sternberg, personal observations). Each site included five plots of 250 $m^{-2}$ (10 × 25 m) with a minimum distance of 10 m between plots. The Mediterranean and semi-arid sites included additional plots with rainfall manipulations that were not considered in the current manuscript.

### Vegetation and seed bank sampling

The sampling of the vegetation was conducted annually at peak biomass – late March in the arid and semi-arid sites, and mid-April in the Mediterranean site, between the growing seasons of 2000/2001 and 2009/2010 (except in 2004/2005). Ten random samples (20 x 20 cm quadrats) of the herbaceous vegetation were taken in the open patches (i.e. patches without shrub cover) in each of the five plots (with a minimum distance of 1m from the plot's edges). Each sample was collected by cutting the vegetation at the ground level and brought to the lab. There, plants were sorted by species, and individuals of each species were counted.

The composition of the seed bank (including both transient and persistent fractions) was estimated by collecting soil cores on an annual basis (2000–2009) in September before the onset of the rainy season. Ten random soil samples were taken from each plot independently of the vegetation samples because the collection of the soil samples is likely to affect the vegetation in that particular sampling area (and vice versa). Soil cores were sampled over an area of 5×5 cm with a soil depth of 5 cm and included surface standing plant litter (c.1-2 cm). Each sample was brought to the lab, thoroughly mixed, and stones and coarse roots were removed. The soil and plant litter was spread in drained plastic trays (12×14 cm, 6.5 cm depth) on a gauze sheet placed on top of a 3-cm-thick layer of perlite. The thickness of the soil layer varied between 0.75 and 1 cm. The trays were irrigated during winter (October-March) in a net-house at the Botanical Garden of Tel Aviv University. Emerging seedlings were identified, counted, and continuously removed until no further emergence was observed a few weeks after the end of irrigation. The overall germinable seed bank from each year was assessed by repeating the germination procedure for each soil sample for three consecutive growing seasons.

Seedling emergence under optimal watering conditions was followed in each soil core for three consecutive years to account for seed dormancy (i.e., seeds that do not germinate after one growing season; Harel, Holzapfel, & Sternberg, 2011). This approach enables a better estimate of the abundance of species with high dormancy fraction. During summer, seed bank trays were naturally dried in the net-house to mimic typical hot, dry field conditions. At the end of the third season, each soil sample was passed through 5- and 0.30-mm sieves, to retrieve non-germinated seeds that were counted under a microscope (80× magnification). Since the number of retrieved non-germinated seeds was very low (<1% of the total number of emerged seedlings) and the procedure very time-consuming, this fraction of the seed bank was not considered in further analyses (see Harel, Holzapfel, & Sternberg, 2011).

The species lists for the three sites are found in the supporting information (Tables S1-S3).

**Statistical Analyses**

Our analysis focuses on the annual species that comprise most of the community in all sites in terms of biomass, abundance, and richness (Tielborger et al., 2014). The seed bank composition was estimated by pooling all seedlings that germinated from each soil core during the three consecutive years of germination. We also performed additional, separate analyses for each year of germination (see Appendix S1 for details). All analyses were based on the Bray-Curtis index (Bray & Curtis, 1957) as a measure of dissimilarity among years and\or between the vegetation and the seed bank. We chose this index which is based on relative abundance data because presence-absence indices are sensitive to variability in the total density (no. of individuals per area) and the spatial scale of the sampling unit (Chase & Knight, 2013). Both the sampling-unit area and total density differed between the seed bank and the vegetation.

One major challenge in temporal analyses is that the dissimilarity in species composition across years can result from sampling errors instead of real temporal variability, especially in heterogenous landscapes (Kalyuzhny et al., 2014). We aimed to minimize the effects of sampling errors among replicates (due to spatial heterogeneity) by aggregating all vegetation and seed bank samples from each year in each site and taking the mean abundance of each species.

To visualize the temporal trends in species composition, we used non-metric multidimensional scaling (NMDS), the most robust ordination method (Minchin, 1987). We used PERMANOVA tests (‘adonis’ function of the ‘vegan’ R package, Oksanen et al., 2019) to test whether the community composition varies among years and between the seed bank and the vegetation. Additionally, we tested for homogeneity of dispersion ('betadisp' function of the 'vegan' R package), one of the assumptions of PERMANOVA tests (Alekseyenko, 2016).

The year-to-year variability was estimated based on the mean distance among all possible pairs of consecutive sampling years. The differences between year-to-year variability in the seed bank and vegetation were compared with a permutation t-test using the 'coin' R package (Hothorn, Winell, Hornik, van de Wiel, & Zeileis, 2019).

To investigate long-term compositional variability we applied a time-lag analysis (Collins, Micheli, & Hartt, 2000) i.e. regressing time-lag (the temporal distance between each pair of years [log transformed]) and compositional dissimilarity (Bray-Curtis). The time-lag analysis is the temporal analog of the commonly used distance-decay approach for spatial analysis of compositional similarity (Nekola & White, 1999). The advantage of the time-lag approach is that it does not require using the first year as a reference point for all other years and allows more accurate estimation because of several replications for each distance class. In this analysis, the slope of the time-lag compositional distance relationship indicates the rate of long-term directional change in composition. We compared the slopes in the vegetation and the seed bank using the method proposed by Nekola & White (1999). This approach, which incorporates the dependence among replications of pairwise distance, was implemented using the 'Simba' R package (Jurasinski & Retzer, 2012)

We investigated whether species' traits can explain differences in composition between the seed bank and the vegetation, as well as differences in temporal trajectories, focusing on seed mass, seed dormancy, and functional group (grasses vs. forbs). These traits were chosen because of their importance for community assembly in the region (DeMalach, Ron, & Kadmon, 2019; Harel et al., 2011). Seed mass data were taken from a previous study in the same sites (Harel, Holzapfel, & Sternberg, 2011) and were available for more than 90% of the individuals sampled. Additionally, a *seed dormancy* index was calculated for each species based on variability in the number of seedlings found in the soil cores during the three consecutive germination years: $\sqrt{\sum_{i=1}^{3}\frac{(i-1)\cdot A_i}{2\mathrm{T}}}$ , where $i$ is the year of germination (not the year of sampling), $A_i$ is the abundance of the species in year $i$ (all soil samples combined) and T is the total abundance of the species (summed over all years). The dormancy index is bounded between zero (when all seeds germinated during the first year) and one (when all seeds germinated during the third year). The square root reduces the skewness of the index resulting from the steep decrease in the number of germinating seeds over the three years. The dormancy index cannot capture dormancy for more than three years, but such long-term dormancy was negligible under net-house conditions (see 'Vegetation and seed bank sampling' section above).

We related species traits and species composition using affinity indices (DeMalach et al., 2019) as a solution for the problem of inflated type I error of the community-weighted mean approach (Miller, Damschen, & Ives, 2018; Peres-Neto, Dray, & ter Braak, 2017). We defined *seed bank affinity* as species' relative abundance in the seed bank compared with the sum of relative abundances in the vegetation and seed bank:

$$seed\ bank\ affinity\ = \frac{A_{seedbank}}{A_{seedbank}+A_{vegetation}}$$

Here, $A_{seedbank}$ and $A_{vegetation}$ represent the relative abundance of the species in the seed bank and the vegetation, respectively (all years pooled together). The seed bank affinity ranges from zero (when a species appears only in the vegetation) to one (appears only in the seed bank). The very rare species that appeared only in the seed bank or in the vegetation were not included in the analyses of seed bank affinity to eliminate the possibility that differences result from low detection rate (see Tables S4–S6 for sample size in the different analyses).

We estimated the effect of the three major traits on seed bank affinity using linear models for species with relative abundance higher than 0.5% to avoid bias caused by rare species with more stochastic occurrences. In the regression, seed mass (mg) was $\log_e$ transformed and the functional group was incorporated as a dummy variable coded one for grasses and zero for forbs. For each regression, we report both the coefficients without transformation (raw estimates) and standardized estimates (when both the explanatory variables and the dependent variable are standardized by subtracting their mean from each observation and then dividing by the standard deviation). Standardized coefficients enable comparison among variables with different units.

## RESULTS

The composition of the seed bank (all three germination years pooled) significantly differed from the composition of the vegetation in the Mediterranean (pseudo-$F_{(1,15)}$ = 2.6, P = 0.016), semi-arid (pseudo-$F_{(1,15)}$ = 3.9, P = 0.003) and arid (pseudo-$F_{(1,15)}$ = 4.3, P < 0.001) sites (Fig. 1, Fig. S1–S3). Heterogeneity of dispersion between the seed bank and the vegetation was found to be significant in the semiarid community (pseudo-$F_{(1,15)}$ =5.1, P = 0.04) and insignificant in the Mediterranean (pseudo-$F_{(1,15)}$ =2.08, P = 0.17) and the arid (pseudo-$F_{(1,15)}$ =0.5, P = 0.48) communities.

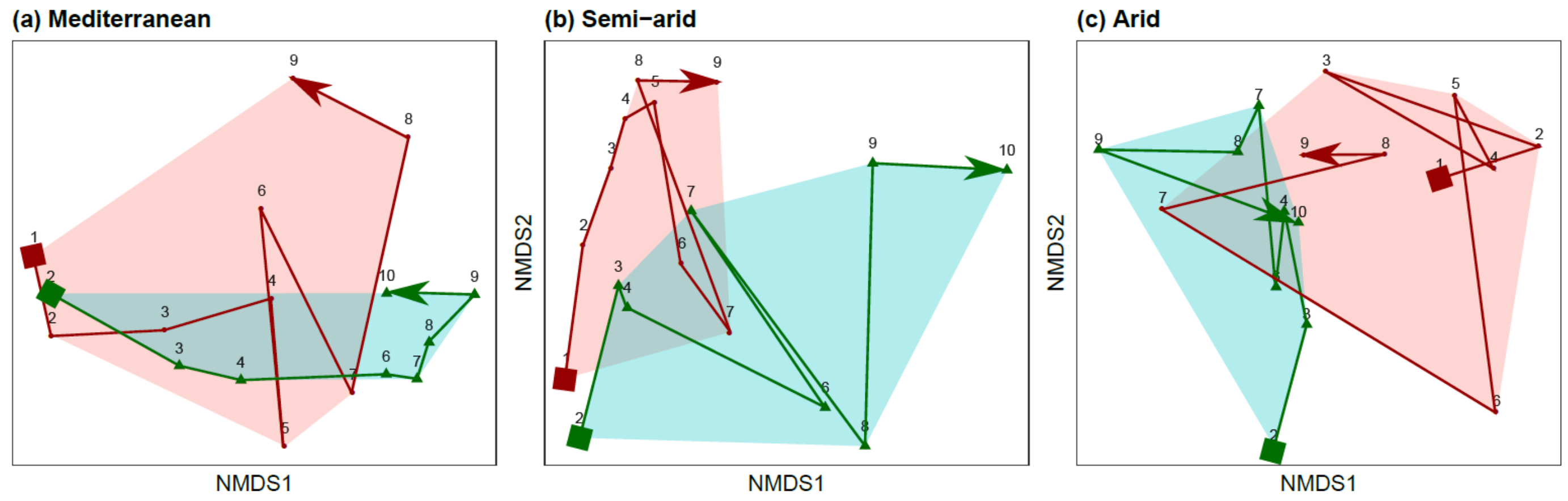

**Figure 1**: Community composition in the seed bank (red circles and polygon) and vegetation (green triangles and polygon) in the three sites represented using non-metric multidimensional scaling (NMDS) based on the Bray–Curtis index. Numbers represent years of sampling (1 – 2001, 2 – 2002, …,10 – 2010). The pink and cyan polygons represent the minimal compositional space occupied by the seed bank and the vegetation. The red and the green arrows represent the temporal trajectories of the community composition of the seed bank and the vegetation. (a) Mediterranean site, stress = 0.15. (b) Semi-arid site, stress = 0.08 (c) Arid site, stress = 0.13.

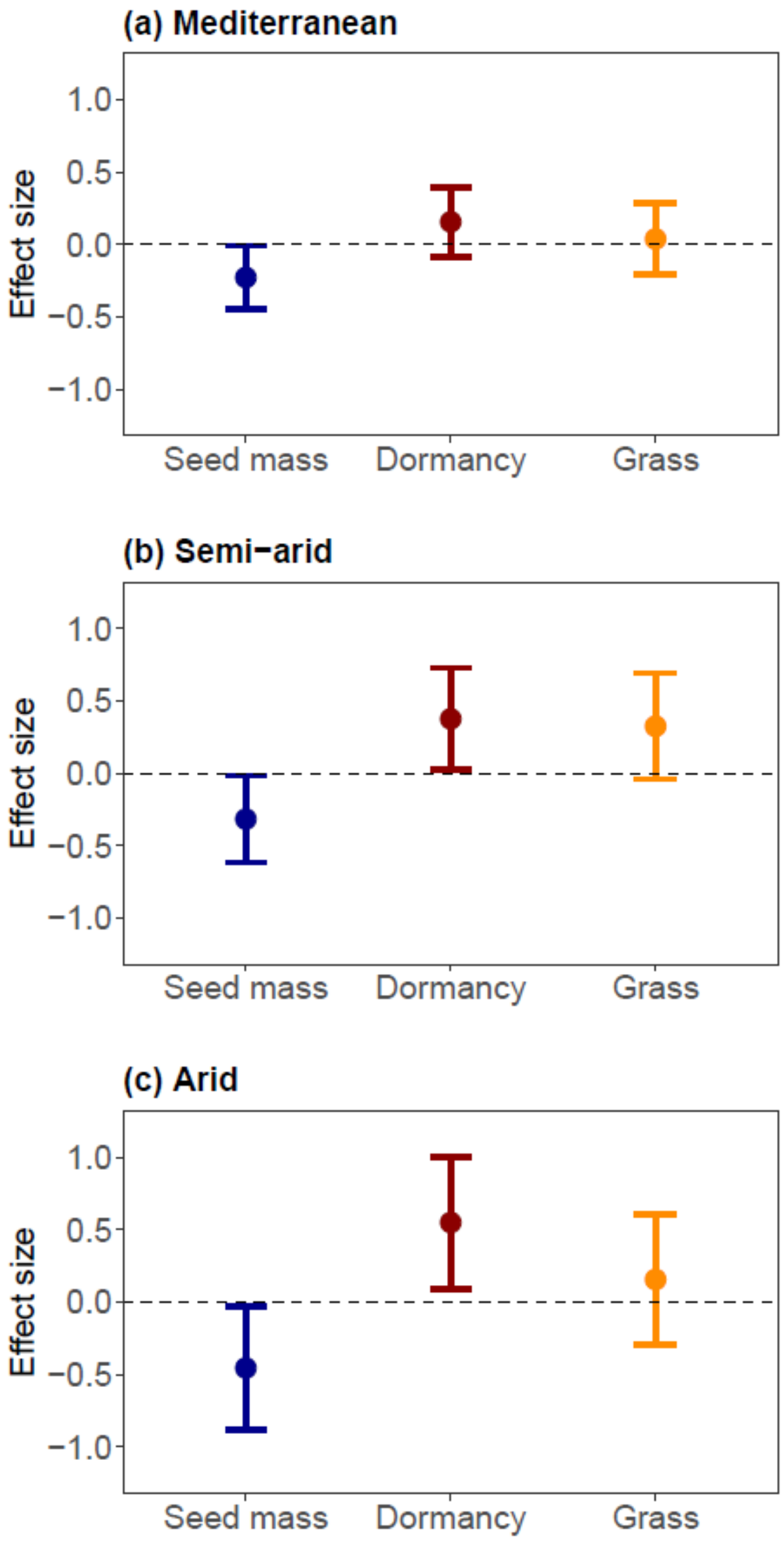

**Figure 2**. The effects of seed mass, seed dormancy index, and functional group membership (coded zero for forbs and one for grasses) on species' seed bank affinity. Effect size (points) represents standardized regression coefficients (see Table S4 for a detailed summary). Species' seed bank affinity (relative abundance in the seed bank compared with that in the vegetation) is negatively affected by seed mass and positively affected by seed dormancy. Error bars represent confidence intervals. The dashed line represents zero effect. $N_{(Mediterranean)} = 80$, $N_{(Semi\text{-}arid)} = 43$, $N_{(Arid)} = 14$.

Species' seed bank affinity (relative abundance in the seed bank compared with the vegetation) was negatively affected by seed mass and positively affected by seed dormancy in the semiarid and arid sites (Fig. 2, Table S4), i.e. small-seeded species and species with higher seed dormancy were more common in the soil seed bank than in the vegetation (but significance levels were marginal in the arid site, Table S4). In the Mediterranean site, seed bank affinity was negatively affected by seed mass and unaffected by dormancy. Plant functional group membership (grasses vs. forbs) did not affect seed bank affinity in any of the sites (Fig. 2, Table S4). Species' seed mass and their dormancy index were not correlated in any of the sites (Fig. S7).

The results did not support our hypothesis that the seed bank undergoes lower year-to-year variability than the vegetation (Fig. 3, blue triangles). Differences in year-to-year variability (dissimilarity between pairs of consecutive years) between the seed bank and the vegetation were not significant in both the semiarid ($Z_{(1,10)}$ = 0.69, P = 0.49) and arid ($Z_{(1,10)}$ = -0.99, P = 0.32) sites. In the Mediterranean site, year-to-year variability was even slightly higher in the seed bank than in the vegetation ($Z_{(1,10)}$, P = 0.013).

Long term directional trends in community composition occurred in the Mediterranean and the semi-arid sites as indicated by the positive relationship between time-lag (temporal distance among years) and compositional distance (Fig. 3a–d, Fig. S8a–d). In contrast, there were no significant relationships between time-lag and compositional distance in the arid site (Fig. 3e–f, Fig. S8e–f).

The rates of long-term changes in the Mediterranean and semi-arid sites (the slopes in Fig. 3) were lower in the seed bank compared with the vegetation (P = 0.013 and P < 0.001, respectively) thereby supporting the hypothesis that the seed bank is more resistant to directional changes than the vegetation. The difference in the slopes was highest in the semi-arid site (about 3.5 times steeper) leading to a larger divergence in composition between the seed bank and vegetation with time (Fig. S9).

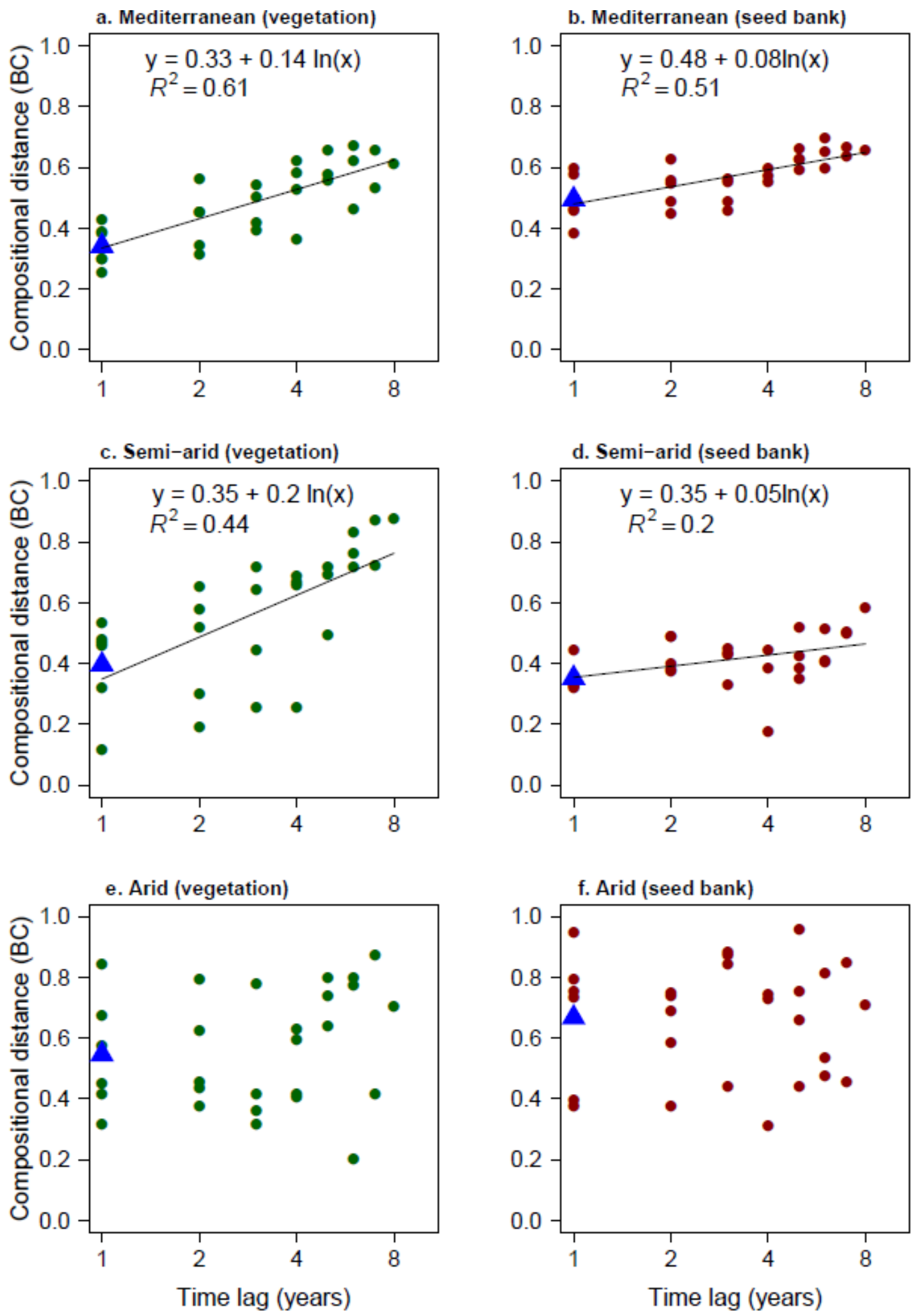

**Figure 3**: Compositional distance (Bray–Curtis index) in the vegetation (left panels) and seed bank (right panels) as a function of time-lag (temporal distance between years of sampling including all possible pairs). The blue triangle represents the mean compositional distance between two consecutive years (year-to-year variability). The slope of the relationship indicates the rate of long-term trends. (a, b) Mediterranean site (c, d) Semi-arid site. (e, f) Arid site. The x-axis has a logarithmic scale. Trendlines were added when the relationship between time-lag and compositional distance was statistically significant ($P<0.05$)

## DISCUSSION

Our findings support the hypothesis that the seed bank is more resistant than the vegetation to long-term compositional shifts in both the Mediterranean and the semi-arid sites. However, the hypothesis of lower year-to-year variability in the seed bank was not supported in any of the sites. Additionally, we demonstrated that the composition of the seed bank differs from the standing vegetation because small-seeded and species with high dormancy fraction are overly represented in the seed bank.

### Differential composition in the seed bank and the vegetation

The PERMANOVA demonstrates major differences in composition between the vegetation and the seed bank in all sites. In the case of the semi-arid site, the results should be treated with caution because the assumption of homogenous dispersion has been violated. Nonetheless, such violation is unlikely to inflate type I error in our study because we had a balanced sample size in the seed bank and the vegetation (see details in Alekseyenko, 2016).

We aimed to explain differences in composition using three major traits: seed mass, seed dormancy, and functional group membership. Seed bank affinity was partially explained by these traits ($R^2$ = 0.07, $R^2$ = 0.18, $R^2$ =0.54, for the Mediterranean, semi-arid and arid sites, respectively), but additional traits could have increased the explanatory power.

Dormant-seeded species were more common in the seed bank than in the vegetation (Fig. 2), as expected for organisms that spend most of their life as seeds and only one growing season as developed plants. Our dormancy index was based on dormancy in net-house conditions with constant irrigation during the growing season which depleted the seed bank after three consecutive growing seasons (see Methods). In natural conditions, however, dormancy could be much longer because of spatial heterogeneity in soil conditions, greater year-to-year variation in rainfall conditions, and other differences between natural and experimental conditions (Kigel, 1995; Thompson, 1987). Therefore, we believe that the association between dormancy and seed bank affinity is even stronger than implied by our analyses.

The finding that small-seeded species were relatively more common in the seed bank (Fig. 2) could be related to several factors. First, small-seeded species often have higher fecundity and are less sensitive to seed predation (Jakobsson & Eriksson, 2000; Lebrija-Trejos, Lobato, & Sternberg, 2011; Petry, Kandlikar, Kraft, Godoy, & Levine, 2018) resulting in higher abundance in the seed bank. At the same time, small-seeded seedlings are more sensitive to abiotic stress (Moles & Westoby, 2004; Muller-Landau, 2010) and size-asymmetric competition (DeMalach et al., 2019) which may reduce their abundance in the vegetation compared with the seed bank. Furthermore, seed size is often correlated with persistence in the soil (Funes, Basconcelo, Díaz, & Cabido, 1999; Thompson, Band, & Hodgson, 1993; Thompson, Bakker, Bekker, & Hodgson, 1998) and with environmental factors controlling germination, particularly light conditions (Kigel, 1995; Thompson, 1987).

In our main analyses, we focused on the total seed bank which included all seeds that germinated during three consecutive watering years after the collection. Comparison of the separate analyses of seeds germinating in the first year only and total seed banks (all years pooled, Fig. 1; see Appendix S1 for details) showed similar patterns of germination (Fig. S4) because the proportion of seeds germinating during the first year was much higher than in the following two years. Moreover, differences between the seed bank and the vegetation were found also when we compared the vegetation to the second and the third years of germination (Fig. S5–S6). In sum, our conclusion that the seed bank undergoes similar year-to-year variability in composition as the vegetation but slower rates of long-term shifts is valid for the total seed bank, as well as the seed bank estimated for each germination year separately (Fig. S10–S12).

All our analyses were based on the Bray-Curtis dissimilarity index which is mostly affected by changes in the relative abundance of common species. The common species in the three sites were found in both the seed bank and the vegetation (Tables S1-S3). However, in each site, some rare species were exclusively found in either the seed bank or the vegetation.  Such patterns could be related to the four times larger area of the vegetation samples or the higher density of the seed bank.

**The role of the seed bank in buffering year-to-year variability**

We used the mean compositional distance among each pair of consecutive years as an indicator of year-to-year variability. This type of short-term variability is often caused by stochastic differences among years in climatic conditions but can also be affected by directional trends (e.g. succession). In our case, we believe that year-to-year variability is mostly related to stochastic variability among years since in the overwhelming majority of cases we did not find a significant overall temporal trend in year-to-year variability (Fig. S13).

Year-to-year variability in the composition of both the seed bank and the vegetation was highest in the arid site which is probably related to the high rainfall variability in this site. However, we caution that despite our aim to minimize alternative sources of variability among sites (see methods), it is difficult to reach generalizations based on three ecosystems.

In contrast with our prediction, year-to-year variability in the seed bank was not lower than in the vegetation. We attribute this finding to species-specific variability in fecundity among years (Venable, 2007) which may lead to high compositional variability in the seed bank. Furthermore, seed bank composition could be affected by variability in dormancy among years due to fluctuations in temperature, soil moisture, granivores, and pathogens (Venable, 2007). Nonetheless, our findings do not imply that the seed bank does not play a role in buffering temporal fluctuations in the vegetation. Even though the relationships between seed bank and vegetation dynamics are highly complex, seed banks can still serve as 'insurance' against population extinctions even when species abundance varies among years (Fischer & Stocklin, 1997).

### The role of the seed bank in buffering long-term shifts

We supported the hypothesis that seed bank composition is more resistant to long-term changes than the vegetation by showing lower rates of directional changes in both the Mediterranean and the semiarid sites (Fig. 3). Directional changes in composition in both sites (Fig. 1, 3) are probably related to the removal of livestock grazing during the establishment of the research sites (Golodets, Kigel, & Sternberg, 2010; Osem, Perevolotsky, & Kigel, 2004; Tielborger et al., 2014). The trend was stronger in the semi-arid site than in the Mediterranean site where past grazing was more intense, while grazing intensity was negligible in the arid site.

### Conclusions

Investigating the drivers of compositional stability is of major importance in times of major climate and land-use changes (Damschen, Harrison, & Grace, 2010; Duprè et al., 2010; Harrison et al., 2015; Komatsu et al., 2019). Several studies have speculated that patterns of vegetation stability are related to seed bank stability. For example, the high drought sensitivity of the vegetation in Californian grasslands was attributed to a depleted seed bank (Harrison, LaForgia, & Latimer, 2018). Furthermore, it has been shown that drought and nitrogen deposition deplete seed banks in several grasslands (Basto et al., 2018, 2015). Conversely, the high stability of Middle-Eastern annual communities to grazing and rainfall changes was attributed to the high resistance of their seed bank to these environmental factors (Sternberg, Gutman, Perevolotsky, & Kigel, 2003; Sternberg et al. 2017; Tielborger et al., 2014). Our results provide empirical support for the above assertion. We have demonstrated that in the Mediterranean and the semi-arid communities, the seed bank undergoes weaker long-term shifts. Therefore, we argue that a better understanding of the buffering role of soil seed banks under climate change will significantly improve our predictions for the future distribution and persistence of annual plant communities.

## ACKNOWLEDGMENTS

We are most grateful to Claus Holzapfel, Hadas Parag, Danny Harel, and Danny Wallach for soil seed bank sampling, and to Irit Konsens for vegetation sampling. We thank two anonymous reviewers and the handling editor for constructive comments on this manuscript. Michael Kalyuzhny provided comments on the temporal analyses. The study was supported by the GLOWA Jordan River project and funded by the German Federal Ministry of Education and Research (BMBF), in collaboration with the Israeli Ministry of Science and Technology (MOST). ND was supported by the Tel Aviv University Postdoctoral Fellowship. The authors declare that they have no conflict of interest.

## AUTHOR CONTRIBUTIONS

MS and JK conceived the research idea within the GLOWA Jordan River project and collected the data. ND developed the seed bank and vegetation comparison, performed the statistical analysis, and wrote the first draft of the paper. All authors substantially contributed to the writing of the manuscript.

## DATA AVAILABILITY STATEMENT

All data will be available on FigShare

## COMPETING INTERESTS STATEMENT

The authors declare no competing financial interests.

## SUPPORTING INFORMATION

### Appendix S1

In the main analyses, seed bank was defined as the total number of seedlings emerging from soil cores, i.e. pooling together the three consecutive years of germination. Additionally, we applied a complementary approach where separate analyses were conducted for each of the three years of germination.

In the arid site, the seed bank composition during the second and third year of germination included a single species – *Filago desertorum*, which emerged in the second germination year of 2001, 2002, 2004, and 2009 sampling years, and the third germination year of 2003 sampling. No seedlings emerged during the third germination year of 2004 and 2006 sampling years. We chose not to exclude years with a single species (these years have identical locations in an NMDS plot and zero distance in time-lag analysis), but years without emerged species were excluded (compositional distance could not be computed).

Overall, the results of these separate analyses were qualitatively similar to the main analysis (Fig S4–S6, S12–S14). Regardless of the type of seed bank analyzed, year-to-year variability of the vegetation was higher compared to the seed bank, and the slope of the time-lag analyses was steeper in the vegetation compared with the seed bank.

**Table S1: Species list for the Mediterranean site sorted by relative abundance. Rank – the rank of relative abundance. RA – mean relative abundance (in the seed bank and the vegetation together). Seed mass – mean seed mass [mg]. Dormancy – dormancy index. Seed bank – occurrence in the seed bank (Y\N). Veg - occurrence in the vegetation (Y\N).**

| Name | Family | Rank | RA | Seed mass | Dormancy | Seed bank | Veg |
|---|---|---|---|---|---|---|---|
| *Brachypodium distachyon* | Gramineae | 1 | 0.1438 | 3.69 | 0.11 | Yes | Yes |
| *Lolium rigidum* | Gramineae | 2 | 0.0676 | 4.59 | 0.31 | Yes | Yes |
| *Plantago afra* | Plantaginaceae | 3 | 0.0663 | 0.68 | 0.38 | Yes | Yes |
| *Catapodium rigidum* | Gramineae | 4 | 0.0522 | 0.194 | 0.37 | Yes | Yes |
| *Convolvulus siculus* | Convolvulaceae | 5 | 0.0496 | NA | 0.33 | Yes | Yes |
| *Avena sterilis* | Gramineae | 6 | 0.0458 | 9.16 | 0.42 | Yes | Yes |
| *Valantia hispida* | Rubiaceae | 7 | 0.0444 | 0.22 | 0.22 | Yes | Yes |
| *Plantago cretica* | Plantaginaceae | 8 | 0.0351 | 1.07 | 0.40 | Yes | Yes |
| *Picris galileae* | Compositae | 9 | 0.0346 | 0.3 | 0.41 | Yes | Yes |
| *Sedum rubens* | Crassulaceae | 10 | 0.0284 | 0.04 | 0.44 | Yes | Yes |
| *Anagallis arvensis* | Primulaceae | 11 | 0.0228 | 0.43 | 0.66 | Yes | Yes |
| *Bromus fasciculatus* | Gramineae | 12 | 0.0217 | 1 | 0.12 | Yes | Yes |
| *Torilis tenella* | Umbelliferae | 13 | 0.0163 | 0.42 | 0.39 | Yes | Yes |
| *Convolvulus pentapetaloides* | Convolvulaceae | 14 | 0.0158 | NA | 0.56 | Yes | Yes |
| *Galium judaicum* | Rubiaceae | 15 | 0.0142 | 0.33 | 0.42 | Yes | Yes |
| *Crepis sancta* | Compositae | 16 | 0.0139 | 0.9 | 0.50 | Yes | Yes |
| *Rhagadiolus stellatus* | Compositae | 17 | 0.0133 | 3.27 | 0.46 | Yes | Yes |
| *Aegilops peregrina* | Gramineae | 18 | 0.0125 | 10.55 | 0.42 | Yes | Yes |
| *Stipa capensis* | Gramineae | 19 | 0.0121 | 2.21 | 0.32 | Yes | Yes |
| *Mercurialis annua* | Euphorbiaceae | 20 | 0.0109 | 0.05 | 0.51 | Yes | Yes |
| *Hedypnois rhagadioloides* | Compositae | 21 | 0.0109 | 2.083 | 0.60 | Yes | Yes |
| *Bromus madritensis* | Gramineae | 22 | 0.0100 | 0 | 0.16 | Yes | Yes |
| *Hordeum spontaneum* | Gramineae | 23 | 0.0099 | 27.3 | 0.23 | Yes | Yes |
| *Galium murale* | Rubiaceae | 24 | 0.0096 | 0.94 | 0.36 | Yes | Yes |
| *Hymenocarpos circinnatus* | Papilionaceae | 25 | 0.0091 | 7.03 | 0.56 | Yes | Yes |
| *Stachys neurocalycina* | Labiatae | 26 | 0.0079 | 0.94 | 0.66 | Yes | Yes |
| *Urospermum picroides* | Compositae | 27 | 0.0075 | 2.71 | 0.51 | Yes | Yes |
| *Trifolium stellatum* | Papilionaceae | 28 | 0.0072 | 2.7 | 0.49 | Yes | Yes |
| *Trifolium pilulare* | Papilionaceae | 29 | 0.0058 | 2.82 | 0.56 | Yes | Yes |
| *Campanula hierosolymitana* | Campanulaceae | 30 | 0.0057 | 0.05 | 0.58 | Yes | Yes |
| *Linum corymbulosum* | Linaceae | 31 | 0.0055 | 4.73 | 0.62 | Yes | Yes |
| *Bromus alopecuros* | Gramineae | 32 | 0.0055 | 1.28 | 0.18 | Yes | Yes |
| *Silene nocturna* | Caryophyllaceae | 33 | 0.0054 | 0.27 | 0.64 | Yes | Yes |
| *Trifolium purpureum* | Papilionaceae | 34 | 0.0052 | 0.73 | 0.59 | Yes | Yes |
| *Parapholis incurva* | Gramineae | 35 | 0.0048 | 1.05 | 0.49 | Yes | Yes |
| *Coronilla scorpioides* | Papilionaceae | 36 | 0.0047 | 20.8 | 0.68 | Yes | Yes |
| *Avena barbata* | Gramineae | 37 | 0.0047 | 5.9 | 0.22 | Yes | No |
| *Filago pyramidata* | Compositae | 38 | 0.0047 | 0.06 | 0.66 | Yes | Yes |
| *Biscutella didyma* | Cruciferae | 39 | 0.0046 | 0.65 | 0.41 | Yes | Yes |
| *Velezia rigida* | Caryophyllaceae | 40 | 0.0046 | 0.26 | 0.58 | Yes | Yes |
| *Alopecurus utriculatus* | Gramineae | 41 | 0.0045 | 1.37 | 0.22 | Yes | Yes |
| *Onobrychis caput galli* | Papilionaceae | 42 | 0.0045 | 14.23 | 0.55 | Yes | Yes |
| *Filago contracta* | Compositae | 43 | 0.0045 | 0.1 | 0.45 | Yes | Yes |
| *Trifolium dasyurum* | Papilionaceae | 44 | 0.0043 | NA | 0.16 | Yes | Yes |
| *Lotus peregrinus* | Papilionaceae | 45 | 0.0043 | 1.91 | 0.53 | Yes | Yes |
| *Torilis leptophylla* | Umbelliferae | 46 | 0.0043 | 2.43 | 0.41 | Yes | Yes |
| *Anthemis pseudocotula* | Compositae | 47 | 0.0042 | 0.41 | 0.47 | Yes | Yes |
| *Pterocephalus plumosus* | Dipsacaceae | 48 | 0.0042 | 2.6 | 0.38 | Yes | Yes |
| *Pimpinella cretica* | Umbelliferae | 49 | 0.0041 | 0.37 | 0.43 | Yes | Yes |
| *Scorpiurus muricatus* | Papilionaceae | 50 | 0.0039 | 1.49 | 0.66 | Yes | Yes |
| *Trifolium campestre* | Papilionaceae | 51 | 0.0038 | 0.53 | 0.47 | Yes | Yes |
| *Medicago monspeliaca* | Papilionaceae | 52 | 0.0037 | 0.72 | 0.54 | Yes | Yes |
| *Filago palaestina* | Compositae | 53 | 0.0036 | 0.08 | 0.68 | Yes | Yes |
| *Trifolium scabrum* | Papilionaceae | 54 | 0.0036 | 1.07 | 0.55 | Yes | Yes |
| *Isatis lusitanica* | Cruciferae | 55 | 0.0033 | 1.7 | 0.38 | Yes | Yes |
| *Helianthemum salicifolium* | Cistaceae | 56 | 0.0033 | 5.43 | 0.20 | Yes | Yes |
| *Crucianella aegyptiaca* | Rubiaceae | 57 | 0.0032 | 0.79 | 0.39 | Yes | Yes |
| *Clypeola jonthlaspi* | Cruciferae | 58 | 0.0031 | 0.21 | 0.37 | Yes | Yes |
| *Erodium malacoides* | Geraniaceae | 59 | 0.0030 | 0.71 | 0.47 | Yes | Yes |
| *Theligonum cynocrambe* | Theligonaceae | 60 | 0.0030 | NA | 0.43 | Yes | Yes |
| *Medicago coronata* | Papilionaceae | 61 | 0.0028 | 0.81 | 0.52 | Yes | Yes |
| *Tordylium trachycarpum* | Umbelliferae | 62 | 0.0027 | NA | 0.36 | Yes | Yes |
| *Scabiosa palaestina* | Dipsacaceae | 63 | 0.0027 | 2.62 | NA | No | Yes |
| *Ziziphora capitata* | Labiatae | 64 | 0.0027 | 0.31 | 0.00 | Yes | Yes |
| *Alyssum strigosum* | Cruciferae | 65 | 0.0025 | NA | NA | No | Yes |
| *Onobrychis squarrosa* | Papilionaceae | 66 | 0.0025 | 16.29 | 0.39 | Yes | Yes |
| *Centaurium tenuiflorum* | Compositae | 67 | 0.0021 | 0.02 | 0.53 | Yes | No |
| *Pterocephalus brevis* | Dipsacaceae | 68 | 0.0020 | 0.6 | 0.42 | Yes | Yes |
| *Medicago rotata* | Papilionaceae | 69 | 0.0019 | 4.61 | 0.57 | Yes | Yes |
| *Lagoecia cuminoides* | Umbelliferae | 70 | 0.0019 | 0.53 | 0.27 | Yes | Yes |
| *Linum strictum* | Linaceae | 71 | 0.0018 | 0.25 | 0.48 | Yes | Yes |
| *Diplotaxis viminea* | Cruciferae | 72 | 0.0018 | 0.2 | 0.50 | Yes | Yes |
| *Thlaspi perfoliatum* | Cruciferae | 73 | 0.0017 | 0.42 | 0.24 | Yes | Yes |
| *Lomelosia palaestina* | Dipsacaceae | 74 | 0.0016 | 0 | 0.34 | Yes | No |
| *Hippocrepis unisiliquosa* | Papilionaceae | 75 | 0.0016 | 3.67 | 0.67 | Yes | Yes |
| *Cephalaria syriaca* | Dipsacaceae | 76 | 0.0013 | 0 | -1.00 | No | Yes |
| *Linum pubescens* | Linaceae | 77 | 0.0013 | 0.75 | 0.59 | Yes | Yes |
| *Chaetosciadium trichospermum* | Umbelliferae | 78 | 0.0013 | 0.8 | 0.63 | Yes | Yes |
| *Psilurus incurvus* | Gramineae | 79 | 0.0013 | 0.05 | 0.44 | Yes | Yes |
| *Misopates orontium* | Scrophulariaceae | 80 | 0.0012 | NA | 0.63 | Yes | Yes |
| *Arenaria leptoclados* | Caryophyllaceae | 81 | 0.0012 | 0.04 | 0.61 | Yes | Yes |
| *Briza maxima* | Gramineae | 82 | 0.0011 | 1.67 | 0.24 | Yes | Yes |
| *Atractylis cancellata* | Compositae | 83 | 0.0011 | 1.44 | 0.46 | Yes | Yes |
| *Medicago orbicularis* | Papilionaceae | 84 | 0.0010 | 5.66 | 0.61 | Yes | Yes |
| *Cephalaria joppensis* | Dipsacaceae | 85 | 0.0010 | 0 | 0.19 | Yes | No |
| *Crupina crupinastrum* | Compositae | 86 | 0.0010 | 22.5 | 0.13 | Yes | Yes |
| *Geropogon hybridus* | Compositae | 87 | 0.0008 | 9.75 | 0.53 | Yes | Yes |
| *Medicago polymorpha* | Papilionaceae | 88 | 0.0008 | 6.09 | 0.77 | Yes | Yes |
| *Alyssum simplex* | Cruciferae | 89 | 0.0008 | 0.65 | 0.22 | Yes | No |
| *Crucianella macrostachya* | Rubiaceae | 90 | 0.0008 | 0.79 | 0.34 | Yes | No |
| *Avena wiestii* | Gramineae | 91 | 0.0007 | 9.14 | 0.86 | Yes | No |
| *Trisetaria macrochaeta* | Gramineae | 92 | 0.0007 | 1.87 | 0.28 | Yes | No |
| *Cicer judaicum* | Papilionaceae | 93 | 0.0007 | 22.34 | 0.64 | Yes | Yes |
| *Rostraria cristata* | Gramineae | 94 | 0.0007 | NA | 0.34 | Yes | Yes |
| *Parapholis filiformis* | Gramineae | 95 | 0.0007 | NA | -1.00 | No | Yes |
| *Cephalaria tenella* | Dipsacaceae | 96 | 0.0007 | NA | 0.00 | Yes | No |
| *Senecio leucanthemifolius* | Compositae | 97 | 0.0006 | 0.25 | 0.71 | Yes | Yes |
| *Sonchus oleraceus* | Compositae | 98 | 0.0006 | 0.18 | 0.00 | Yes | Yes |
| *Helianthemum aegyptiacum* | Cistaceae | 99 | 0.0006 | NA | 0.00 | No | Yes |
| *Galium setaceum* | Rubiaceae | 100 | 0.0005 | 0.09 | 0.52 | Yes | Yes |
| *Vicia palaestina* | Papilionaceae | 101 | 0.0005 | 26.04 | 0.63 | No | Yes |
| *Bromus japonicus* | Gramineae | 102 | 0.0005 | NA | 0.20 | Yes | Yes |
| *Erophila praecox* | Cruciferae | 103 | 0.0005 | 0.03 | 0.61 | No | Yes |
| *Crithopsis delileana* | Gramineae | 104 | 0.0005 | 4.1 | 0.25 | Yes | Yes |
| *Geranium rotundifolium* | Geraniaceae | 105 | 0.0005 | 2.6 | 0.53 | Yes | Yes |
| *Euphorbia chamaepeplus* | Euphorbiaceae | 106 | 0.0004 | 1.17 | 0.51 | Yes | Yes |
| *Callipeltis cucullaria* | Rubiaceae | 107 | 0.0004 | 0.1 | 0.36 | Yes | Yes |
| *Bromus lanceolatus* | Gramineae | 108 | 0.0004 | NA | 0.18 | Yes | Yes |
| *Silene colorata* | Caryophyllaceae | 109 | 0.0004 | NA | 0.52 | Yes | No |
| *Daucus durieua* | Umbelliferae | 110 | 0.0004 | 1.67 | 0.69 | Yes | No |
| *Erodium gruinum* | Geraniaceae | 111 | 0.0004 | 57.33 | 0.45 | Yes | Yes |
| *Catananche lutea* | Compositae | 112 | 0.0004 | 2.55 | 0.41 | No | Yes |
| *Ononis mollis* | Papilionaceae | 113 | 0.0004 | NA | 0.29 | Yes | Yes |
| *Ononis ornithopodioides* | Papilionaceae | 114 | 0.0004 | 1.67 | 0.61 | Yes | Yes |
| *Euphorbia oxyodonta* | Euphorbiaceae | 115 | 0.0004 | NA | 0.63 | Yes | Yes |
| *Galium cassium* | Rubiaceae | 116 | 0.0004 | NA | 0.26 | Yes | No |
| *Euphorbia exigua* | Euphorbiaceae | 117 | 0.0003 | 0.15 | 0.64 | Yes | Yes |
| *Erodium moschatum* | Geraniaceae | 118 | 0.0003 | 5.47 | 0.22 | Yes | No |
| *Anchusa aegyptiaca* | Boraginaceae | 119 | 0.0003 | 5.97 | 0.48 | Yes | Yes |
| *Lomelosia porphyroneura* | Dipsacaceae | 120 | 0.0003 | NA | 0.82 | Yes | No |
| *Telmissa microcarpa* | Crassulaceae | 121 | 0.0003 | NA | 0.38 | Yes | No |
| *Astragalus asterias* | Papilionaceae | 122 | 0.0003 | NA | 0.34 | Yes | Yes |
| *Parietaria lusitanica* | Urticaceae | 123 | 0.0003 | NA | 0.76 | Yes | Yes |
| *Crassula alata* | Crassulaceae | 124 | 0.0003 | NA | 0.49 | Yes | No |
| *Lathyrus blepharicarpos* | Papilionaceae | 125 | 0.0003 | NA | 0.58 | Yes | Yes |
| *Althaea hirsuta* | Malvaceae | 126 | 0.0003 | NA | 0.29 | Yes | Yes |
| *Reichardia tingitana* | Compositae | 127 | 0.0002 | 1.13 | 0.52 | Yes | Yes |
| *Veronica cymbalaria* | Scrophulariaceae | 128 | 0.0002 | NA | 0.00 | Yes | No |
| *Sherardia arvensis* | Rubiaceae | 129 | 0.0002 | NA | 0.56 | Yes | No |
| *Trifolium tomentosum* | Papilionaceae | 130 | 0.0002 | NA | NA | No | Yes |
| *Trigonella hierosolymitana* | Papilionaceae | 131 | 0.0002 | NA | 0.58 | Yes | Yes |
| *Crepis aspera* | Compositae | 132 | 0.0002 | 0.19 | 0.85 | Yes | Yes |
| *Medicago tuberculata* | Papilionaceae | 133 | 0.0002 | NA | 0.87 | Yes | Yes |
| *Linum nodiflorum* | Linaceae | 134 | 0.0002 | 8.61 | 0.43 | Yes | Yes |
| *Minuartia decipiens* | Caryophyllaceae | 135 | 0.0002 | NA | NA | No | Yes |
| *Minuartia hybrida* | Caryophyllaceae | 136 | 0.0001 | 0.11 | 0.45 | Yes | Yes |
| *Cuscuta spp* | Convolvulaceae | 137 | 0.0001 | NA | NA | No | Yes |
| *Astragalus epiglottis* | Papilionaceae | 138 | 0.0001 | 1.61 | 0.71 | No | Yes |
| *Medicago truncatula* | Papilionaceae | 139 | 0.0001 | 4.93 | 0.75 | Yes | No |
| *Trigonella spinosa* | Papilionaceae | 140 | 0.0001 | NA | 0.58 | Yes | Yes |
| *Cichorium endivia* | Compositae | 141 | 0.0001 | 1.07 | 0.62 | Yes | Yes |
| *Notobasis syriaca* | Compositae | 142 | 0.0001 | NA | 0.00 | No | Yes |
| *Trifolium resupinatum* | Papilionaceae | 143 | 0.0001 | NA | 0.50 | Yes | No |
| *Euphorbia helioscopia* | Euphorbiaceae | 144 | 0.0001 | NA | NA | No | Yes |
| *Trifolium cherleri* | Papilionaceae | 145 | 0.0001 | NA | 0.00 | Yes | Yes |
| *Factorovskya aschersoniana* | Papilionaceae | 146 | 0.0001 | NA | 0.00 | Yes | No |
| *Aegilops kotschyi* | Gramineae | 147 | 0.0001 | NA | 0.35 | Yes | No |
| *Geranium molle* | Geraniaceae | 148 | 0.0001 | NA | NA | No | Yes |
| *Trifolium arguntum* | Papilionaceae | 149 | 0.0001 | NA | NA | No | Yes |
| *Ononis sicula* | Papilionaceae | 150 | 0.0001 | 1.3 | NA | No | Yes |
| *Erodium subintegrifolium* | Geraniaceae | 151 | 0.0001 | NA | 0.00 | Yes | No |
| *Minuartia mediterranea* | Caryophyllaceae | 153 | 0.0001 | NA | 0.00 | Yes | No |
| *Silene decipiens* | Caryophyllaceae | 153 | 0.0001 | NA | 0.31 | Yes | No |
| *Valerianella vesicaria* | Valerianaceae | 154 | 0.0001 | 3.48 | 0.49 | No | Yes |
| *Diplotaxis harra* | Cruciferae | 156 | 0.0001 | NA | 0.29 | Yes | No |
| *Ononis viscosa* | Papilionaceae | 156 | 0.0001 | NA | 0.00 | Yes | No |
| *Minuartia picta* | Caryophyllaceae | 157 | 0.0001 | NA | 0.39 | Yes | No |
| *Pisum sativum* | Papilionaceae | 158 | 0.0001 | NA | NA | No | Yes |
| *Alopecurus myosuroides* | Gramineae | 160 | 4E-05 | NA | 0.29 | Yes | No |
| *Carthamus glaucus* | Compositae | 160 | 4E-05 | NA | 0.58 | Yes | No |
| *Astragalus tribuloides* | Papilionaceae | 161 | 4E-05 | 4.63 | 0.00 | No | Yes |
| *Trifolium clusii* | Papilionaceae | 162 | 4E-05 | 1.02 | NA | No | Yes |
| *Erodium cicutarium* | Geraniaceae | 164 | 3E-05 | 1.25 | 0.43 | Yes | No |
| *Hypochaeris achyrophorus* | Compositae | 164 | 3E-05 | NA | 0.41 | Yes | No |
| *Silene alexandrina* | Caryophyllaceae | 165 | 3E-05 | NA | NA | No | Yes |
| *Euphorbia peplus* | Euphorbiaceae | 166 | 3E-05 | 1.7 | 0.60 | Yes | No |
| *Onobrychis crista galli* | Papilionaceae | 167 | 3E-05 | NA | 0.60 | No | Yes |
| *Scandix verna* | Umbelliferae | 168 | 3E-05 | NA | NA | No | Yes |
| *Calendula arvensis* | Compositae | 169 | 3E-05 | 1.15 | 0.50 | No | Yes |
| *Vicia sativa* | Papilionaceae | 170 | 2E-05 | NA | NA | No | Yes |
| *Silene aegyptica* | Caryophyllaceae | 172 | 2E-05 | NA | NA | No | Yes |
| *Trifolium clypeatum* | Papilionaceae | 172 | 2E-05 | 5 | 0.00 | No | Yes |
| *Helianthemum lasiocarpum* | Cistaceae | 173 | 1E-05 | NA | NA | No | Yes |
| *Filago desertorum* | Compositae | 174 | 1E-05 | 0.03 | 0.46 | No | Yes |
| *Centaurea cyanoides* | Compositae | 177 | 8E-06 | NA | NA | No | Yes |
| *Crepis senecioides* | Compositae | 177 | 8E-06 | 0.09 | 0.29 | No | Yes |
| *Medicago minima* | Papilionaceae | 177 | 8E-06 | NA | NA | No | Yes |
| *Vulpia myuros* | Gramineae | 177 | 8E-06 | NA | NA | No | Yes |

**Table S2: Species list for the semi-arid site sorted by relative abundance. Rank – the rank of relative abundance. RA – mean relative abundance (in the seed bank and the vegetation together). Seed mass – mean seed mass [mg]. Dormancy – dormancy index. Seed bank – occurrence in the seed bank (Y\N). Veg - occurrence in the vegetation (Y\N).**

| Name | Family | Rank | RA | Seed mass | Dormancy | Seed bank | Veg |
|---|---|---|---|---|---|---|---|
| *Trisetaria macrochaeta* | Gramineae | 1 | 0.3638 | 1.87 | 0.28 | Yes | Yes |
| *Crithopsis delileana* | Gramineae | 2 | 0.1367 | 4.1 | 0.25 | Yes | Yes |
| *Filago contracta* | Compositae | 3 | 0.0984 | 0.1 | 0.45 | Yes | Yes |
| *Aegilops peregrina* | Gramineae | 4 | 0.0417 | 10.55 | 0.42 | Yes | Yes |
| *Atractylis cancellata* | Compositae | 5 | 0.0335 | 1.44 | 0.46 | Yes | Yes |
| *Carrichtera annua* | Cruciferae | 6 | 0.0319 | 1.35 | 0.54 | Yes | Yes |
| *Cichorium endivia* | Compositae | 7 | 0.0289 | 1.07 | 0.62 | Yes | Yes |
| *Stipa capensis* | Gramineae | 8 | 0.0256 | 2.21 | 0.32 | Yes | Yes |
| *Brachypodium distachyon* | Gramineae | 9 | 0.0252 | 3.69 | 0.11 | Yes | Yes |
| *Sedum rubens* | Crassulaceae | 10 | 0.0224 | 0.04 | 0.44 | Yes | Yes |
| *Catapodium rigidum* | Gramineae | 11 | 0.0221 | 0.194 | 0.37 | Yes | Yes |
| *Anagallis arvensis* | Primulaceae | 12 | 0.0215 | 0.43 | 0.66 | Yes | Yes |
| *Erophila minima* | Cruciferae | 13 | 0.021 | NA | NA | No | Yes |
| *Onobrychis crista galli* | Papilionaceae | 14 | 0.0121 | NA | 0.60 | Yes | Yes |
| *Rostraria cristata* | Gramineae | 15 | 0.009 | NA | 0.34 | Yes | Yes |
| *Hedypnois rhagadioloides* | Compositae | 16 | 0.0076 | 2.083 | 0.60 | Yes | Yes |
| *Psilurus incurvus* | Gramineae | 17 | 0.0063 | 0.05 | 0.44 | Yes | Yes |
| *Plantago cretica* | Plantaginaceae | 18 | 0.0063 | 1.07 | 0.40 | Yes | Yes |
| *Parapholis incurva* | Gramineae | 19 | 0.0056 | 1.05 | 0.49 | Yes | Yes |
| *Plantago coronopus* | Plantaginaceae | 20 | 0.0052 | 0.38 | 0.38 | Yes | Yes |
| *Aegilops kotschyi* | Gramineae | 21 | 0.0049 | NA | 0.35 | Yes | No |
| *Lolium rigidum* | Gramineae | 22 | 0.0047 | 4.59 | 0.31 | Yes | Yes |
| *Hippocrepis unisiliquosa* | Papilionaceae | 23 | 0.0045 | 3.67 | 0.67 | Yes | Yes |
| *Filago desertorum* | Compositae | 24 | 0.004 | 0.03 | 0.46 | Yes | Yes |
| *Hymenocarpos circinnatus* | Papilionaceae | 25 | 0.0038 | 7.03 | 0.56 | Yes | Yes |
| *Minuartia picta* | Caryophyllaceae | 26 | 0.0036 | NA | 0.39 | Yes | Yes |
| *Filago palaestina* | Compositae | 27 | 0.0035 | 0.08 | 0.68 | Yes | Yes |
| *Medicago monspeliaca* | Papilionaceae | 28 | 0.0033 | 0.72 | 0.54 | Yes | Yes |
| *Crassula alata* | Crassulaceae | 29 | 0.003 | NA | 0.49 | Yes | No |
| *Linum pubescens* | Linaceae | 30 | 0.0028 | 0.75 | 0.59 | No | Yes |
| *Daucus durieua* | Umbelliferae | 31 | 0.0023 | 1.67 | 0.69 | Yes | Yes |
| *Bromus fasciculatus* | Gramineae | 32 | 0.0021 | 1 | 0.12 | Yes | Yes |
| *Parapholis filiformis* | Gramineae | 33 | 0.002 | NA | NA | No | Yes |
| *Euphorbia chamaepeplus* | Euphorbiaceae | 34 | 0.0018 | 1.17 | 0.51 | Yes | Yes |
| *Onobrychis squarrosa* | Papilionaceae | 35 | 0.0016 | 16.29 | 0.39 | Yes | Yes |
| *Urospermum picroides* | Compositae | 36 | 0.0016 | 2.71 | 0.51 | Yes | No |
| *Linum strictum* | Linaceae | 37 | 0.0014 | 0.25 | 0.48 | Yes | Yes |
| *Sonchus oleraceus* | Compositae | 38 | 0.0012 | 0.18 | 0.00 | Yes | No |
| *Crepis sancta* | Compositae | 39 | 0.0012 | 0.9 | 0.50 | Yes | Yes |
| *Diplotaxis viminea* | Cruciferae | 40 | 0.0011 | 0.2 | 0.50 | Yes | No |
| *Valantia hispida* | Rubiaceae | 41 | 0.0009 | 0.22 | 0.22 | Yes | Yes |
| *Herniaria hirsuta* | Caryophyllaceae | 42 | 0.0009 | 0.1 | 0.27 | Yes | Yes |
| *Bromus madritensis* | Gramineae | 43 | 0.0008 | NA | 0.16 | Yes | No |
| *Filago pyramidata* | Compositae | 44 | 0.0008 | 0.06 | 0.66 | Yes | Yes |
| *Torilis tenella* | Umbelliferae | 45 | 0.0007 | 0.42 | 0.39 | Yes | Yes |
| *Picris galileae* | Compositae | 46 | 0.0007 | 0.3 | 0.41 | Yes | Yes |
| *Linum corymbulosum* | Linaceae | 47 | 0.0006 | 4.73 | 0.62 | No | Yes |
| *Campanula hierosolymitana* | Campanulaceae | 48 | 0.0006 | 0.05 | 0.58 | Yes | Yes |
| *Bromus alopecuros* | Gramineae | 49 | 0.0006 | 1.28 | 0.18 | Yes | Yes |
| *Bromus lanceolatus* | Gramineae | 50 | 0.0006 | NA | 0.18 | Yes | Yes |
| *Callipeltis cucullaria* | Rubiaceae | 51 | 0.0006 | 0.1 | 0.36 | Yes | Yes |
| *Hordeum spontaneum* | Gramineae | 52 | 0.0006 | 27.3 | 0.23 | Yes | No |
| *Rhagadiolus stellatus* | Compositae | 53 | 0.0005 | 3.27 | 0.46 | Yes | Yes |
| *Silene nocturna* | Caryophyllaceae | 54 | 0.0005 | 0.27 | 0.64 | Yes | Yes |
| *Biscutella didyma* | Cruciferae | 55 | 0.0005 | 0.65 | 0.41 | Yes | Yes |
| *Minuartia hybrida* | Caryophyllaceae | 56 | 0.0005 | 0.11 | 0.45 | Yes | Yes |
| *Erodium malacoides* | Geraniaceae | 57 | 0.0005 | 0.71 | 0.47 | Yes | Yes |
| *Mercurialis annua* | Euphorbiaceae | 58 | 0.0005 | 0.05 | 0.51 | Yes | No |
| *Anthemis pseudocotula* | Compositae | 59 | 0.0005 | 0.41 | 0.47 | Yes | Yes |
| *Pterocephalus brevis* | Dipsacaceae | 60 | 0.0004 | 0.6 | 0.42 | Yes | Yes |
| *Crepis aspera* | Compositae | 61 | 0.0004 | 0.19 | 0.85 | Yes | No |
| *Avena barbata* | Gramineae | 62 | 0.0004 | 5.9 | 0.22 | Yes | No |
| *Velezia rigida* | Caryophyllaceae | 63 | 0.0004 | 0.26 | 0.58 | Yes | Yes |
| *Chaetosciadium trichospermum* | Umbelliferae | 64 | 0.0004 | 0.8 | 0.63 | Yes | No |
| *Scabiosa palaestina* | Dipsacaceae | 65 | 0.0004 | 2.62 | NA | No | Yes |
| *Isatis lusitanica* | Cruciferae | 66.5 | 0.0003 | 1.7 | 0.38 | Yes | No |
| *Trifolium purpureum* | Papilionaceae | 66.5 | 0.0003 | 0.73 | 0.59 | Yes | No |
| *Avena sterilis* | Gramineae | 68 | 0.0003 | 9.16 | 0.42 | Yes | Yes |
| *Adonis dentata* | Ranunculaceae | 69 | 0.0003 | 0 | NA | No | Yes |
| *Galium judaicum* | Rubiaceae | 70 | 0.0003 | 0.33 | 0.42 | Yes | Yes |
| *Reichardia tingitana* | Compositae | 71 | 0.0003 | 1.13 | 0.52 | Yes | Yes |
| *Helianthemum salicifolium* | Cistaceae | 72 | 0.0003 | 5.43 | 0.20 | No | Yes |
| *Clypeola jonthlaspi* | Cruciferae | 73 | 0.0003 | 0.21 | 0.37 | Yes | Yes |
| *Crepis senecioides* | Compositae | 74 | 0.0003 | 0.09 | 0.29 | Yes | Yes |
| *Crucianella aegyptiaca* | Rubiaceae | 75 | 0.0002 | 0.79 | 0.39 | Yes | Yes |
| *Anchusa aegyptiaca* | Boraginaceae | 76 | 0.0002 | 5.97 | 0.48 | Yes | Yes |
| *Pterocephalus plumosus* | Dipsacaceae | 77 | 0.0002 | 2.6 | 0.38 | Yes | No |
| *Theligonum cynocrambe* | Theligonaceae | 78 | 0.0002 | NA | 0.43 | Yes | No |
| *Senecio leucanthemifolius* | Compositae | 79 | 0.0002 | 0.25 | 0.71 | Yes | No |
| *Trifolium scabrum* | Papilionaceae | 80 | 0.0002 | 1.07 | 0.55 | Yes | Yes |
| *Galium murale* | Rubiaceae | 81.5 | 0.0001 | 0.94 | 0.36 | Yes | No |
| *Tordylium trachycarpum* | Umbelliferae | 81.5 | 0.0001 | NA | 0.36 | Yes | No |
| *Pteranthus dichotomus* | Caryophyllaceae | 83 | 0.0001 | NA | NA | No | Yes |
| *Catananche lutea* | Compositae | 84 | 0.0001 | 2.55 | 0.41 | Yes | Yes |
| *Schismus arabicus* | Gramineae | 85 | 0.0001 | 0.05 | 0.19 | No | Yes |
| *Geranium rotundifolium* | Geraniaceae | 86.5 | 0.0001 | 2.6 | 0.53 | Yes | No |
| *Micropus supinus* | Compositae | 86.5 | 0.0001 | 37.31 | 0.69 | Yes | No |
| *Factorovskya aschersoniana* | Papilionaceae | 88 | 1E-04 | NA | 0.00 | No | Yes |
| *Lotus peregrinus* | Papilionaceae | 89 | 1E-04 | 1.91 | 0.53 | Yes | No |
| *Erodium gruinum* | Geraniaceae | 90 | 9E-05 | 57.33 | 0.45 | No | Yes |
| *Lagoecia cuminoides* | Umbelliferae | 91.5 | 9E-05 | 0.53 | 0.27 | Yes | No |
| *Onobrychis caput galli* | Papilionaceae | 91.5 | 9E-05 | 14.23 | 0.55 | Yes | No |
| *Bupleurum lancifolium* | Umbelliferae | 93 | 9E-05 | 4.57 | 0.71 | No | Yes |
| *Astragalus asterias* | Papilionaceae | 94 | 8E-05 | NA | 0.34 | No | Yes |
| *Salvia viridis* | Labiatae | 95 | 8E-05 | 2.4 | 0.20 | No | Yes |
| *Mericarpaea ciliata* | Rubiaceae | 96 | 6E-05 | NA | 0.00 | Yes | Yes |
| *Centaurea hyalolepis* | Compositae | 99 | 6E-05 | NA | 0.00 | Yes | No |
| *Convolvulus siculus* | Convolvulaceae | 99 | 6E-05 | NA | 0.33 | Yes | No |
| *Crucianella macrostachya* | Rubiaceae | 99 | 6E-05 | 0.79 | 0.34 | Yes | No |
| *Erodium laciniatum* | Geraniaceae | 99 | 6E-05 | 1.25 | 0.35 | Yes | No |
| *Erodium moschatum* | Geraniaceae | 99 | 6E-05 | 5.47 | 0.22 | Yes | No |
| *Plantago bellardii* | Plantaginaceae | 102 | 5E-05 | NA | NA | No | Yes |
| *Scabiosa prolifera* | Dipsacaceae | 103 | 5E-05 | NA | NA | No | Yes |
| *Plantago afra* | Plantaginaceae | 104 | 4E-05 | 0.68 | 0.38 | Yes | No |
| *Valerianella vesicaria* | Valerianaceae | 105 | 4E-05 | 3.48 | 0.49 | No | Yes |
| *Scorpiurus muricatus* | Papilionaceae | 106 | 4E-05 | 1.49 | 0.66 | No | Yes |
| *Carthamus glaucus* | Compositae | 107 | 4E-05 | NA | 0.58 | No | Yes |
| *Calendula arvensis* | Compositae | 108 | 4E-05 | 1.15 | 0.50 | No | Yes |
| *Ononis mollis* | Papilionaceae | 109 | 3E-05 | NA | 0.29 | No | Yes |
| *Silene aegyptica* | Caryophyllaceae | 110 | 3E-05 | NA | NA | No | Yes |
| *Minuartia decipiens* | Caryophyllaceae | 111 | 2E-05 | NA | NA | No | Yes |
| *Plantago ovata* | Plantaginaceae | 113 | 2E-05 | NA | 0.00 | No | Yes |
| *Tripodion tetraphyllum* | Papilionaceae | 113 | 2E-05 | NA | 0.71 | No | Yes |
| *Silene alexandrina* | Caryophyllaceae | 114 | 1E-05 | NA | NA | No | Yes |
| *Astragalus epiglottis* | Papilionaceae | 117 | 1E-05 | 1.61 | 0.71 | No | Yes |
| *Astragalus tribuloides* | Papilionaceae | 117 | 1E-05 | 4.63 | 0.00 | No | Yes |
| *Convolvulus pentapetaloides* | Convolvulaceae | 117 | 1E-05 | NA | 0.56 | No | Yes |
| *Geropogon hybridus* | Compositae | 117 | 1E-05 | 9.75 | 0.53 | No | Yes |
| *Thlaspi perfoliatum* | Cruciferae | 117 | 1E-05 | 0.42 | 0.24 | No | Yes |
| *Alyssum strigosum* | Cruciferae | 121 | 1E-05 | NA | NA | No | Yes |
| *Galium setaceum* | Rubiaceae | 121 | 1E-05 | 0.09 | 0.52 | No | Yes |
| *Vulpia muralis* | Gramineae | 121 | 1E-05 | NA | NA | No | Yes |
| *Bromus tectorum* | Gramineae | 124 | 7E-06 | NA | NA | No | Yes |
| *Carthamus tenuis* | Compositae | 124 | 7E-06 | NA | NA | No | Yes |
| *Bromus japonicus* | Gramineae | 127 | 5E-06 | NA | 0.20 | No | Yes |
| *Ononis ornithopodioides* | Papilionaceae | 127 | 5E-06 | 1.67 | 0.61 | No | Yes |
| *Papaver hybridum* | Papaveraceae | 127 | 5E-06 | NA | NA | No | Yes |
| *Trifolium tomentosum* | Papilionaceae | 127 | 5E-06 | NA | NA | No | Yes |

**Table S3: Species list for the arid site sorted by relative abundance. Rank – the rank of relative abundance. RA – mean relative abundance (in the seed bank and the vegetation together). Seed mass – mean seed mass [mg]. Dormancy – dormancy index. Seed bank – occurrence in the seed bank (Y\N). Veg - occurrence in the vegetation (Y\N).**

| Name | Family | Rank | RA | Seed mass | Dormancy | Seed bank | Veg |
|---|---|---|---|---|---|---|---|
| *Crepis sancta* | Compositae | 1 | 0.254 | 0.9 | 0.50 | Yes | Yes |
| *Malva aegyptia* | Malvaceae | 2 | 0.1867 | 1.4 | 0.00 | Yes | Yes |
| *Herniaria hirsuta* | Caryophyllaceae | 3 | 0.1211 | 0.1 | 0.27 | Yes | Yes |
| *Plantago bellardii* | Plantaginaceae | 4 | 0.1188 | NA | NA | Yes | Yes |
| *Trisetaria macrochaeta* | Gramineae | 5 | 0.0888 | 1.87 | 0.28 | Yes | Yes |
| *Lappula spinocarpos* | Boraginaceae | 6 | 0.0387 | 7.62 | NA | Yes | Yes |
| *Astragalus tribuloides* | Papilionaceae | 7 | 0.0251 | 4.63 | 0.00 | Yes | Yes |
| *Schismus arabicus* | Gramineae | 8 | 0.0213 | 0.05 | 0.19 | Yes | Yes |
| *Urospermum picroides* | Compositae | 9 | 0.02 | 2.71 | 0.51 | Yes | Yes |
| *Lolium rigidum* | Gramineae | 10 | 0.0152 | 4.59 | 0.31 | Yes | Yes |
| *Plantago coronopus* | Plantaginaceae | 11 | 0.011 | 0.38 | 0.38 | Yes | Yes |
| *Catapodium rigidum* | Gramineae | 12 | 0.0079 | 0.194 | 0.37 | Yes | No |
| *Gymnarrhena micrantha* | Compositae | 13 | 0.0074 | NA | 0.50 | Yes | Yes |
| *Carrichtera annua* | Cruciferae | 14 | 0.0065 | 1.35 | 0.54 | Yes | No |
| *Crepis aspera* | Compositae | 15 | 0.0061 | 0.19 | 0.85 | Yes | No |
| *Avena wiestii* | Gramineae | 16 | 0.0058 | 9.14 | 0.86 | Yes | No |
| *Picris longirostris* | Compositae | 17 | 0.0055 | 0.3 | 0.41 | Yes | No |
| *Biscutella didyma* | Cruciferae | 18 | 0.0048 | 0.65 | 0.41 | Yes | No |
| *Cichorium endivia* | Compositae | 19 | 0.0042 | 1.07 | 0.62 | Yes | No |
| *Galium judaicum* | Rubiaceae | 20 | 0.0039 | 0.33 | 0.42 | Yes | No |
| *Plantago afra* | Plantaginaceae | 21 | 0.0037 | 0.68 | 0.38 | No | Yes |
| *Reichardia tingitana* | Compositae | 22 | 0.0034 | 1.13 | 0.52 | No | Yes |
| *Bromus fasciculatus* | Gramineae | 23 | 0.0034 | 1 | 0.12 | Yes | Yes |
| *Cuscuta spp* | Convolvulaceae | 24 | 0.0023 | NA | NA | Yes | Yes |
| *Hippocrepis unisiliquosa* | Papilionaceae | 25.5 | 0.0022 | 3.67 | 0.67 | Yes | No |
| *Filago contracta* | Compositae | 25.5 | 0.0022 | 0.1 | 0.45 | Yes | No |
| *Erodium cicutarium* | Geraniaceae | 27 | 0.0022 | 1.25 | 0.43 | Yes | No |
| *Erucaria microcarpa* | Cruciferae | 28 | 0.002 | 0.45 | 0.00 | Yes | No |
| *Gastrocotyle hispida* | Boraginaceae | 29 | 0.0019 | NA | NA | Yes | Yes |
| *Euphorbia chamaepeplus* | Euphorbiaceae | 30 | 0.0018 | 1.17 | 0.51 | No | Yes |
| *Erodium touchyanum* | Geraniaceae | 31.5 | 0.0016 | NA | 0.00 | Yes | No |
| *Filago palaestina* | Compositae | 31.5 | 0.0016 | 0.08 | 0.68 | Yes | No |
| *Helianthemum salicifolium* | Cistaceae | 33 | 0.0016 | 5.43 | 0.20 | Yes | Yes |
| *Anthemis melampodina* | Compositae | 34 | 0.0015 | 0.41 | 0.00 | No | Yes |
| *Erodium laciniatum* | Geraniaceae | 35 | 0.0014 | 1.25 | 0.35 | No | Yes |
| *Silene decipiens* | Caryophyllaceae | 36 | 0.0014 | NA | 0.31 | No | Yes |
| *Anagallis arvensis* | Primulaceae | 37 | 0.0013 | 0.43 | 0.66 | Yes | Yes |
| *Helianthemum lasiocarpum* | Cistaceae | 38.5 | 0.0013 | NA | NA | Yes | No |
| *Spergula fallax* | Caryophyllaceae | 38.5 | 0.0013 | NA | NA | Yes | No |
| *Avena sterilis* | Gramineae | 41.5 | 0.0011 | 9.16 | 0.42 | Yes | No |
| *Filago desertorum* | Compositae | 41.5 | 0.0011 | 0.03 | 0.46 | Yes | No |
| *Stipa capensis* | Gramineae | 41.5 | 0.0011 | 2.21 | 0.32 | Yes | No |
| *Isatis lusitanica* | Cruciferae | 41.5 | 0.0011 | 1.7 | 0.38 | Yes | No |
| *Leontodon laciniatus* | Compositae | 44 | 0.0009 | 0.24 | NA | No | Yes |
| *Valantia hispida* | Rubiaceae | 45 | 0.0009 | 0.22 | 0.22 | No | Yes |
| *Filago pyramidata* | Compositae | 46 | 0.0008 | 0.06 | 0.66 | No | Yes |
| *Sonchus oleraceus* | Compositae | 47 | 0.0007 | 0.18 | 0.00 | Yes | No |
| *Arenaria leptoclados* | Caryophyllaceae | 48 | 0.0003 | 0.04 | 0.61 | No | Yes |
| *Plantago ovata* | Plantaginaceae | 49 | 0.0003 | NA | 0.00 | No | Yes |
| *Bupleurum lancifolium* | Umbelliferae | 50.5 | 0.0002 | 4.57 | 0.71 | No | Yes |
| *Minuartia hybrida* | Caryophyllaceae | 50.5 | 0.0002 | 0.11 | 0.45 | No | Yes |
| *Spergularia diandra* | Caryophyllaceae | 52 | 0.0002 | 0.09 | 0.00 | No | Yes |
| *Diplotaxis harra* | Cruciferae | 53 | 0.0001 | NA | 0.29 | No | Yes |

**Table S4: Results of linear models of species' seed bank affinity as a function of their seed mass ($log_e$ transformed [mg]), seed dormancy index, and functional group membership (0 – forbs, 1 – grasses).** Standardized estimates for the regression coefficients are calculated by standardizing both the explanatory and the dependent variables to enable comparison among variables varying in units (see methods).

| | Mediterranean | | | | Semi-arid | | | | Arid | | | |
|---|---|---|---|---|---|---|---|---|---|---|---|---|
| | (raw) estimate | Std. estimate | Std. error | p | (raw) estimate | Std. estimate | Std. error | p | (raw) estimate | Std. estimate | Std. error | p |
| (intercept) | 0.37 | 0.00 | 0.13 | 0.212 | 0.13 | 0.00 | 0.20 | 0.463 | 0.20 | 0.00 | 0.14 | 0.184 |
| Seed mass | -0.03 | **-0.23** | 0.02 | **0.05** | -0.06 | **-0.32** | 0.03 | **0.046** | -0.08 | **-0.45** | 0.04 | **0.063** |
| Dormancy | 0.21 | 0.16 | 0.17 | 0.21 | 0.75 | **0.38** | 0.36 | **0.043** | 0.74 | **0.55** | 0.31 | **0.040** |
| Grass | 0.14 | 0.04 | 0.07 | 0.76 | 0.22 | 0.33 | 0.13 | 0.09 | 0.10 | 0.16 | 0.15 | 0.512 |
| N | 80 | | | | 43 | | | | 14 | | | |
| $R^2$ | **0.07** | | | | **0.18** | | | | **0.54** | | | |

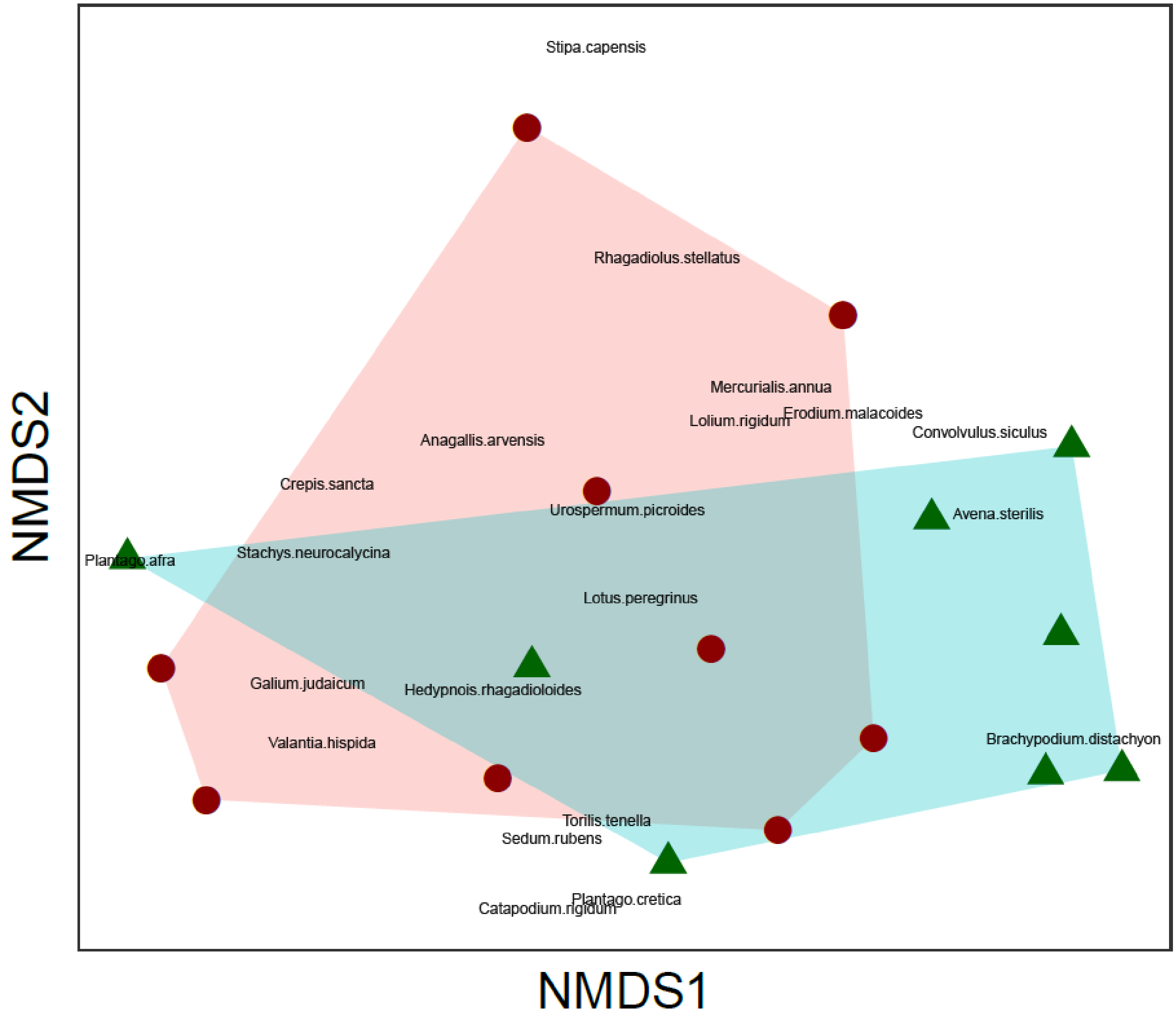

**Fig. S1**: Community composition in the seed bank (brown circles) and vegetation (green triangles) in the Mediterranean site represented using non-metric multidimensional scaling (NMDS) of the 30 most abundant species (instead of all species as in the main text) based on Bray–Curtis dissimilarity. The pink and cyan polygons represent the minimal compositional space occupied by the seed bank and the vegetation. Stress = 0.11.

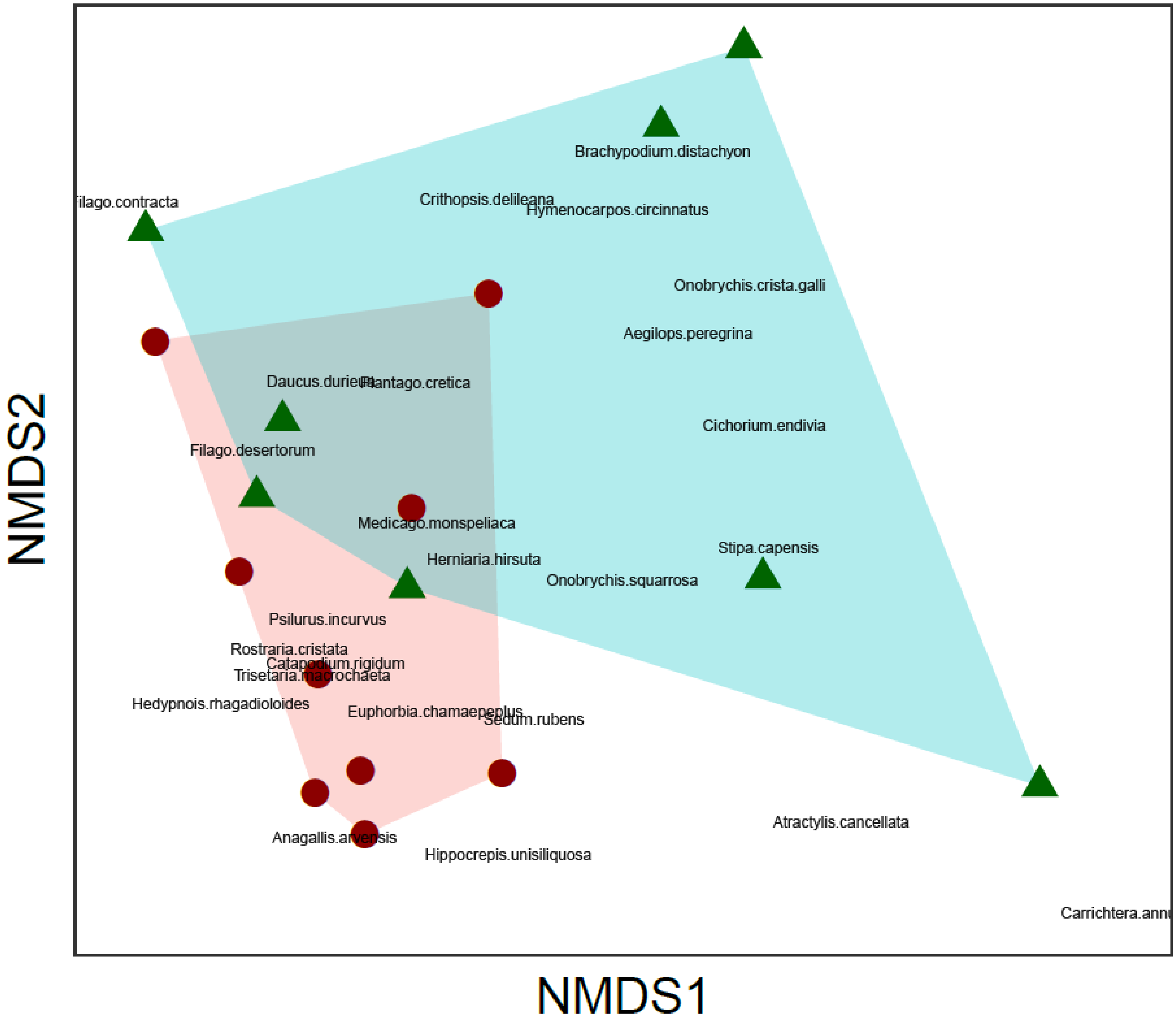

**Fig. S2:** Community composition in the seed bank (brown circles) and vegetation (green triangles) in the Semi-arid site represented using non-metric multidimensional scaling (NMDS) of the 30 most abundant species (instead of all species as in the main text) based on Bray–Curtis dissimilarity. The pink and cyan polygons represent the minimal compositional space occupied by the seed bank and the vegetation. Stress = 0.06

## (c) Arid

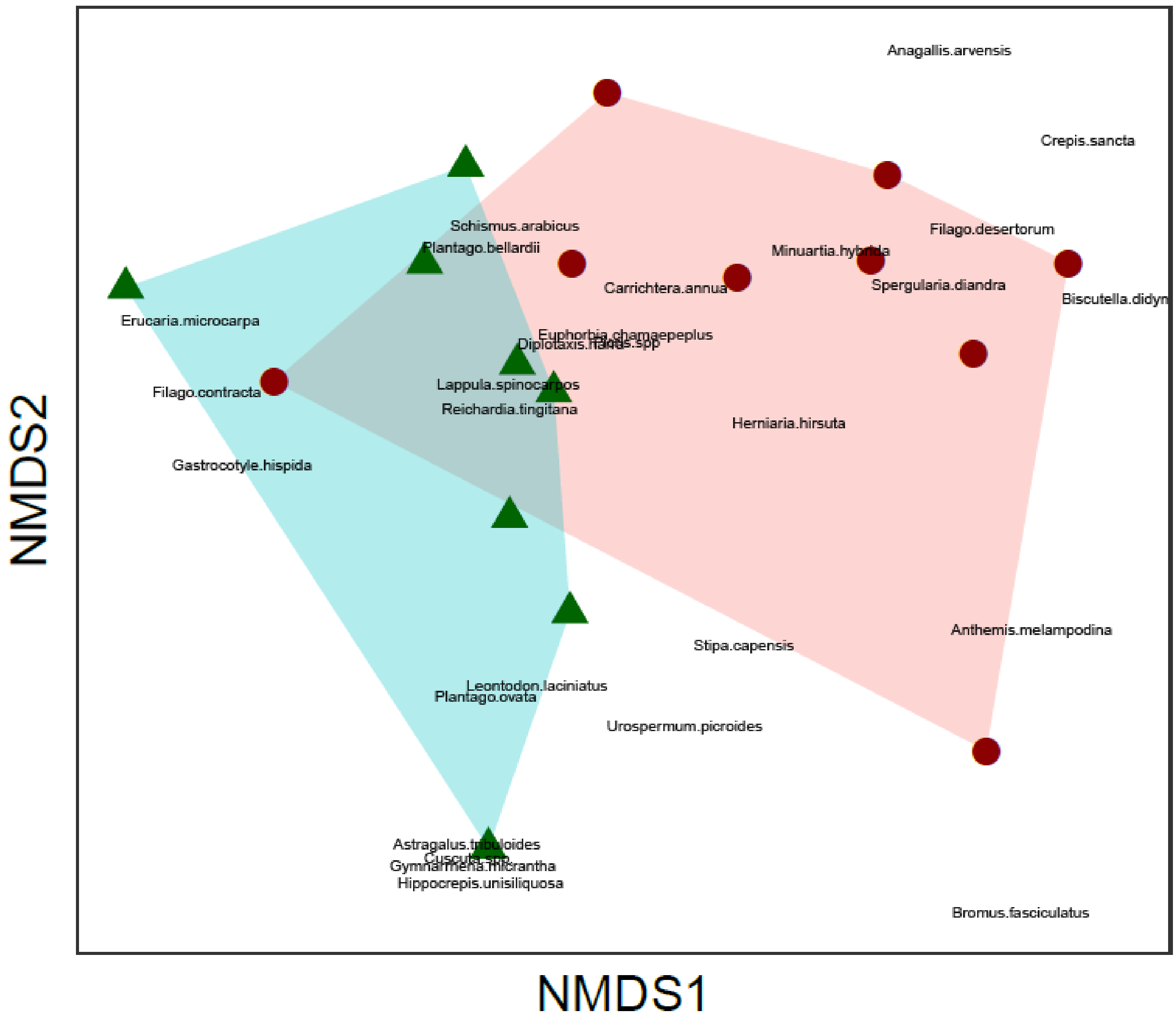

**Fig. S3:** Community composition in the seed bank (brown circles) and vegetation (green triangles) in the Arid site represented using non-metric multidimensional scaling (NMDS) of the 30 most abundant species (instead of all species as in the main text) based on Bray–Curtis dissimilarity. Names of all species used in the analysis are shown. The pink and cyan polygons represent the minimal compositional space occupied by the seed bank and the vegetation. Stress = 0.13

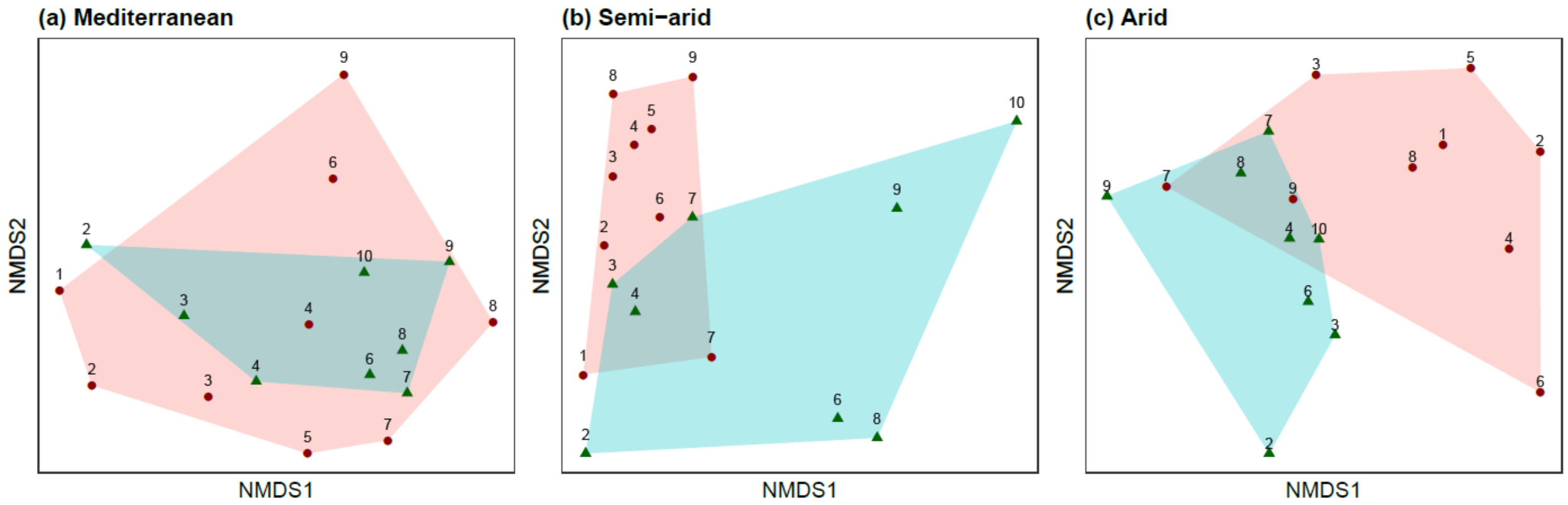

**Fig. S4:** Community composition in the seed bank from *the first germination season* (brown circles) and vegetation (green triangles) in the three sites represented using non-metric multidimensional scaling (NMDS) based on Bray–Curtis dissimilarity. (a) Mediterranean site, stress=0.15, Pseudo-$F_{PERMANOVA(1,15)}$ = 2.27, $P_{PERMANOVA}$ = 0.03, Pseudo-$F_{DISPERSION(1,15)}$ = 2.7, $P_{DISPERSION}$ = 0.12. (b) Semi-arid site, stress=0.09, Pseudo-$F_{PERMANOVA(1,15)}$ = 4.75, $P_{PERMANOVA}$ = 0.004, Pseudo-$F_{DISPERSION(1,15)}$ = 7.4, $P_{DISPERSION}$ = 0.01. (c) Arid site, stress=0.12, Pseudo-$F_{PERMANOVA(1,15)}$ = 3.8, $P_{PERMANOVA}$ = 0.009, Pseudo-$F_{DISPERSION(1,15)}$ = 0.67, $P_{DISPERSION}$ = 0.44.

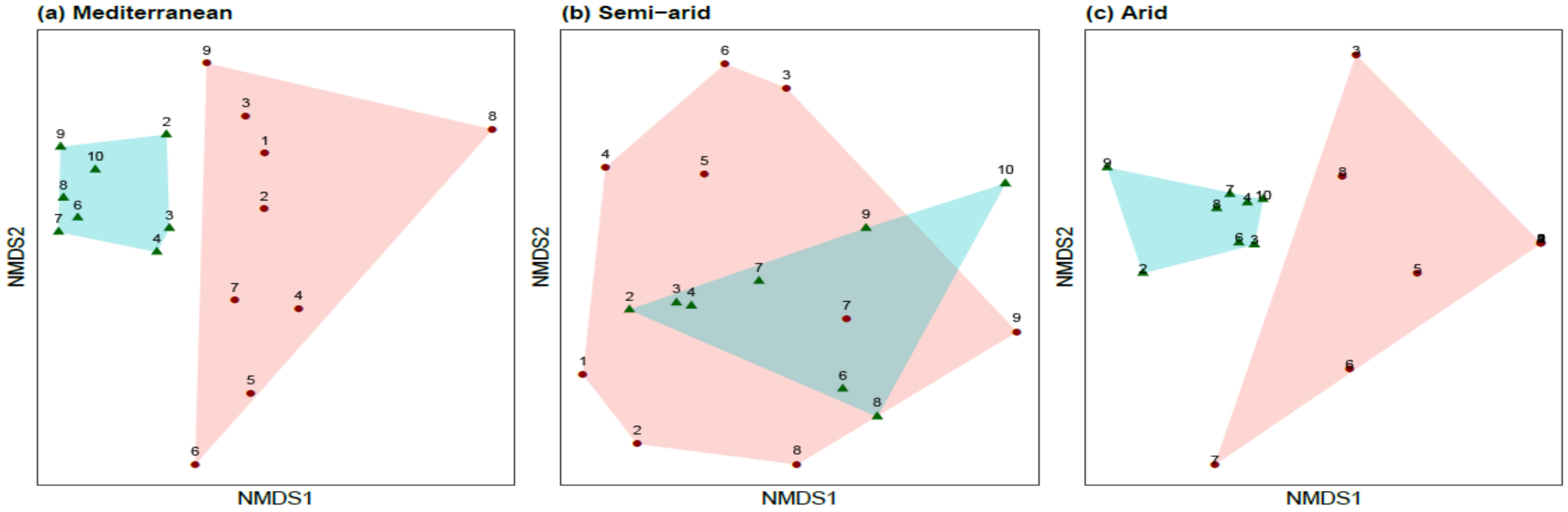

**Fig. S5:** Community composition in the seed bank from *the second germination season* (brown circles) and vegetation (green triangles) in the three sites represented using non-metric multidimensional scaling (NMDS) based on Bray–Curtis dissimilarity. (a) Mediterranean site, stress=0.14, Pseudo-$F_{PERMANOVA(1,15)}$ = 4.68, $P_{PERMANOVA}$ < 0.001, Pseudo-$F_{DISPERSION(1,15)}$ = 11.825, $P_{DISPERSION}$ = 0.003.(b) Semi-arid site, stress=0.15, Pseudo-$F_{PERMANOVA(1,15)}$ = 1.97, $P_{PERMANOVA}$ = 0.04, Pseudo-$F_{DISPERSION(1,15)}$ = 4.6, $P_{DISPERSION}$ = 0.04. (c) Arid site, stress=0.09, Pseudo-$F_{PERMANOVA(1,15)}$ = 6.8, $P_{PERMANOVA}$ < 0.001, Pseudo-$F_{DISPERSION(1,15)}$ = 0.46, $P_{DISPERSION}$ = 0.54.

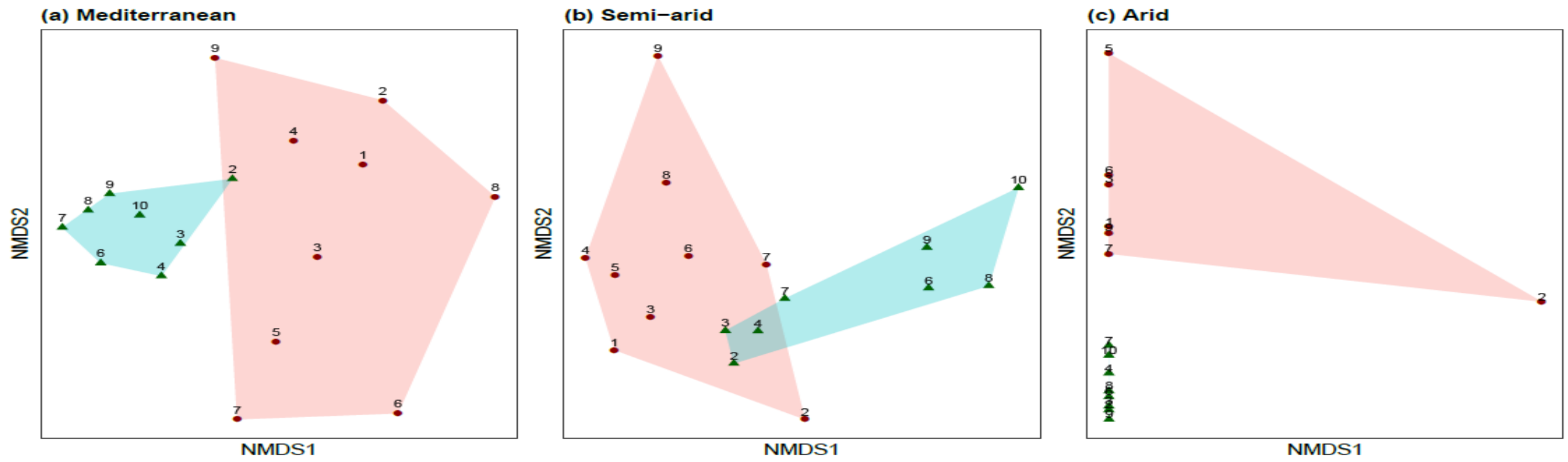

**Fig. S6:** Community composition in the seed bank from *the third germination season* (brown circles) and vegetation (green triangles) in the three sites represented using non-metric multidimensional scaling (NMDS) based on Bray–Curtis dissimilarity. (a) Mediterranean site, stress=0.14, Pseudo-$F_{PERMANOVA(1,15)}$ = 5.5, $P_{PERMANOVA}$ < 0.001, Pseudo-$F_{DISPERSION(1,15)}$ = 18.9, $P_{DISPERSION}$ = 0.001.(b) Semi-arid site, stress=0.12, Pseudo-$F_{PERMANOVA(1,15)}$ = 4.02, $P_{PERMANOVA}$ = 0.001, Pseudo-$F_{DISPERSION(1,15)}$ = 0.75, $P_{DISPERSION}$ = 0.4 (c) Arid site, stress=$10^{-5}$, Pseudo-$F_{PERMANOVA(1,15)}$ = 5.6, $P_{PERMANOVA}$ < 0.001, Pseudo-$F_{DISPERSION(1,15)}$ = 4.44, $P_{DISPERSION}$ = 0.05.

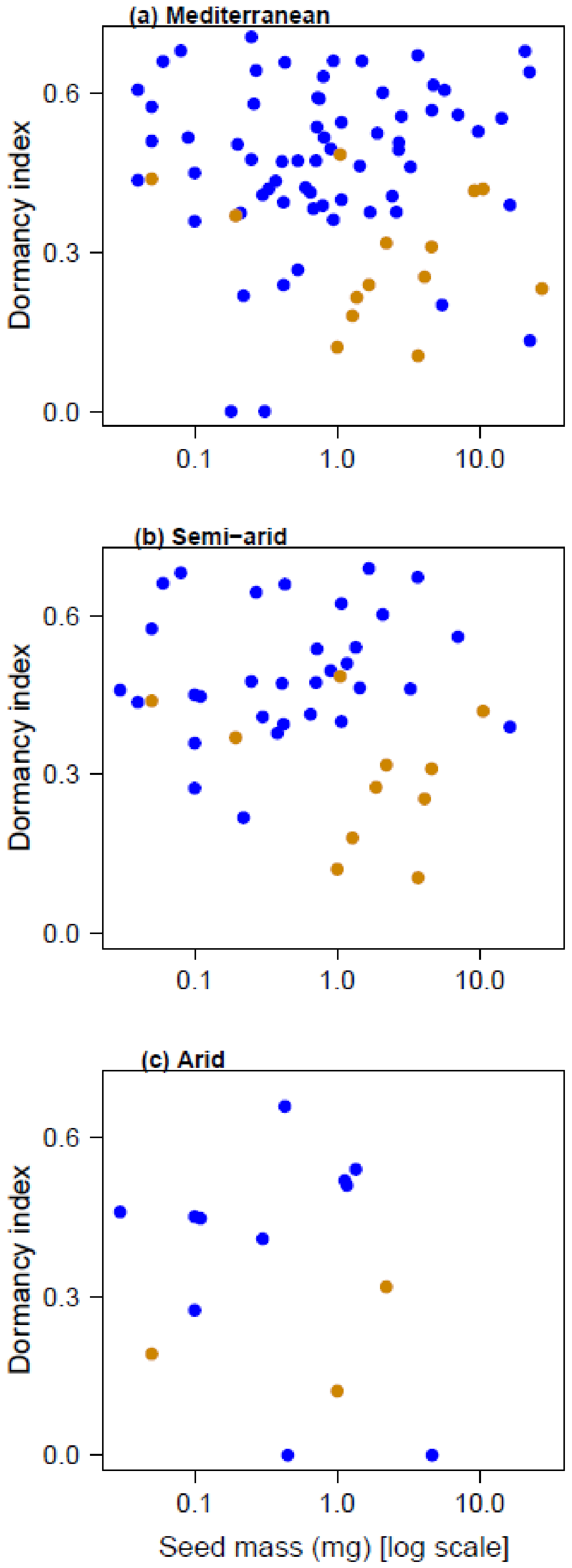

**Fig. S7.** Relationships between seed mass and seed dormancy. Orange points represent grass species while blue circles represent forb species. (a) Mediterranean site (N=80) (b) Semi-arid site (N=43) (c) Arid site (N=14). The relationships were not significant for any of the sites. The x-axes in the left panels are in logarithmic scale.

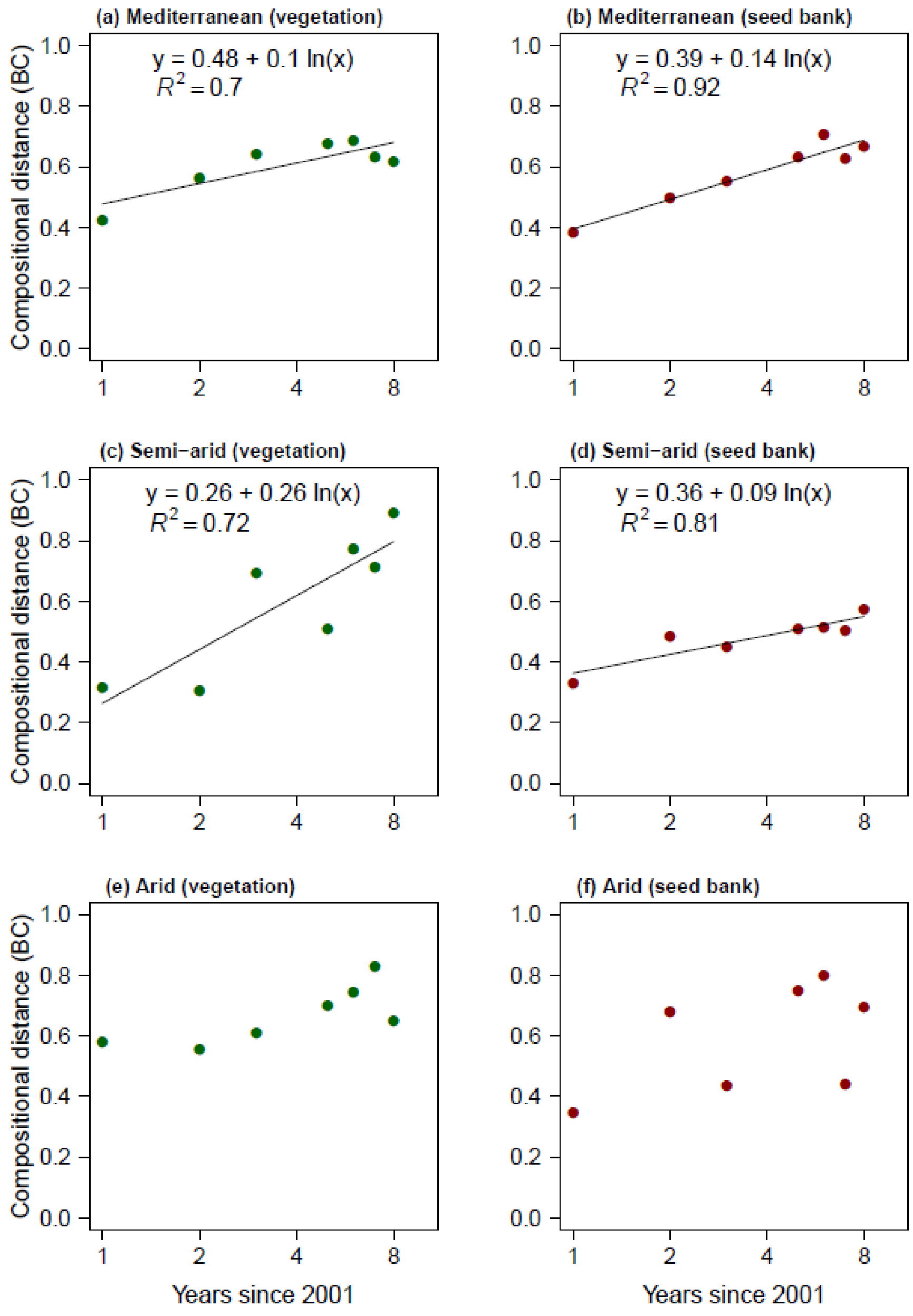

**Fig S8:** The relationship between the temporal distance from the first growing season (2001/2002) and the compositional distance (Bray–Curtis index) for the years 2002/2003–2009/2010. Note the log scale of the x-axis. A trend line appears when there is a significant linear trend (P<0.05).

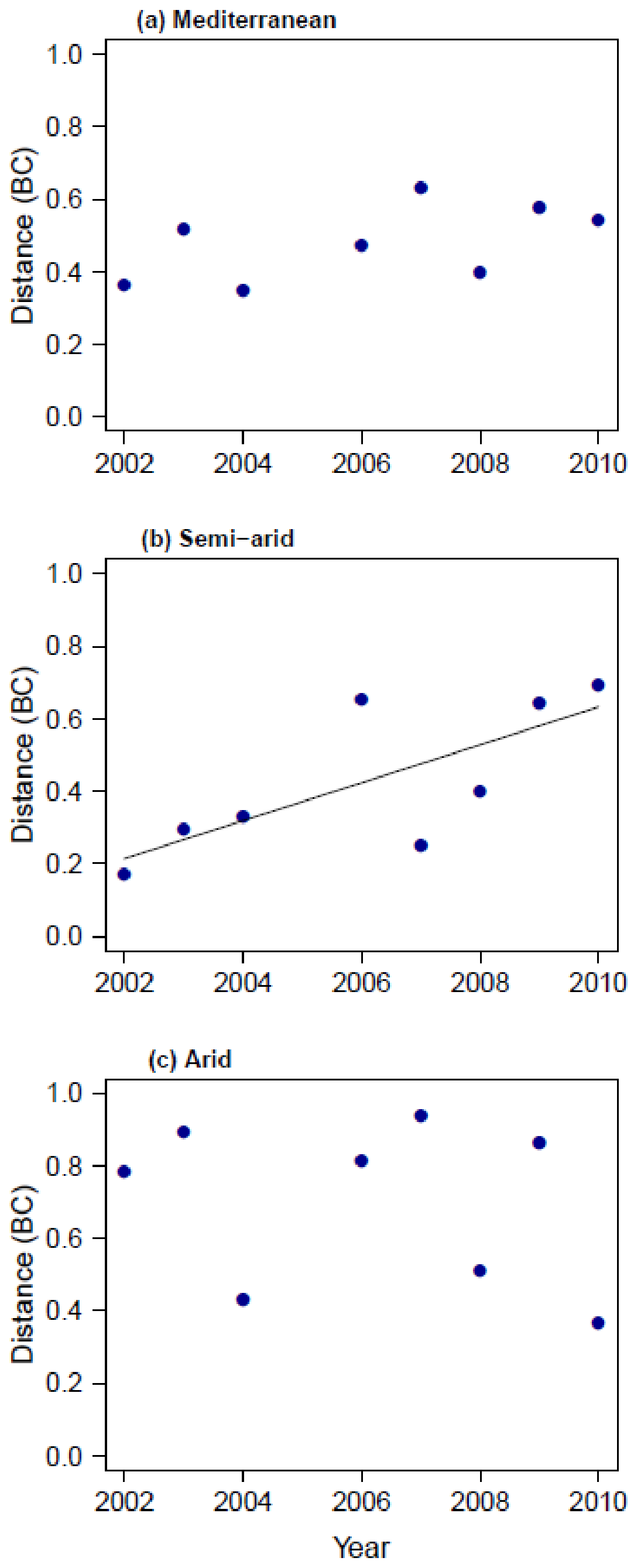

**Fig. S9.** Temporal dynamics of Bray–Curtis distance between the seed bank and the vegetation in the three sites. Each point represents the distance between the vegetation collected during March\April and the seed bank that was collected before the vegetation during September of the previous calendar year. A trend line appears when there is a significant linear trend ($P<0.05$).

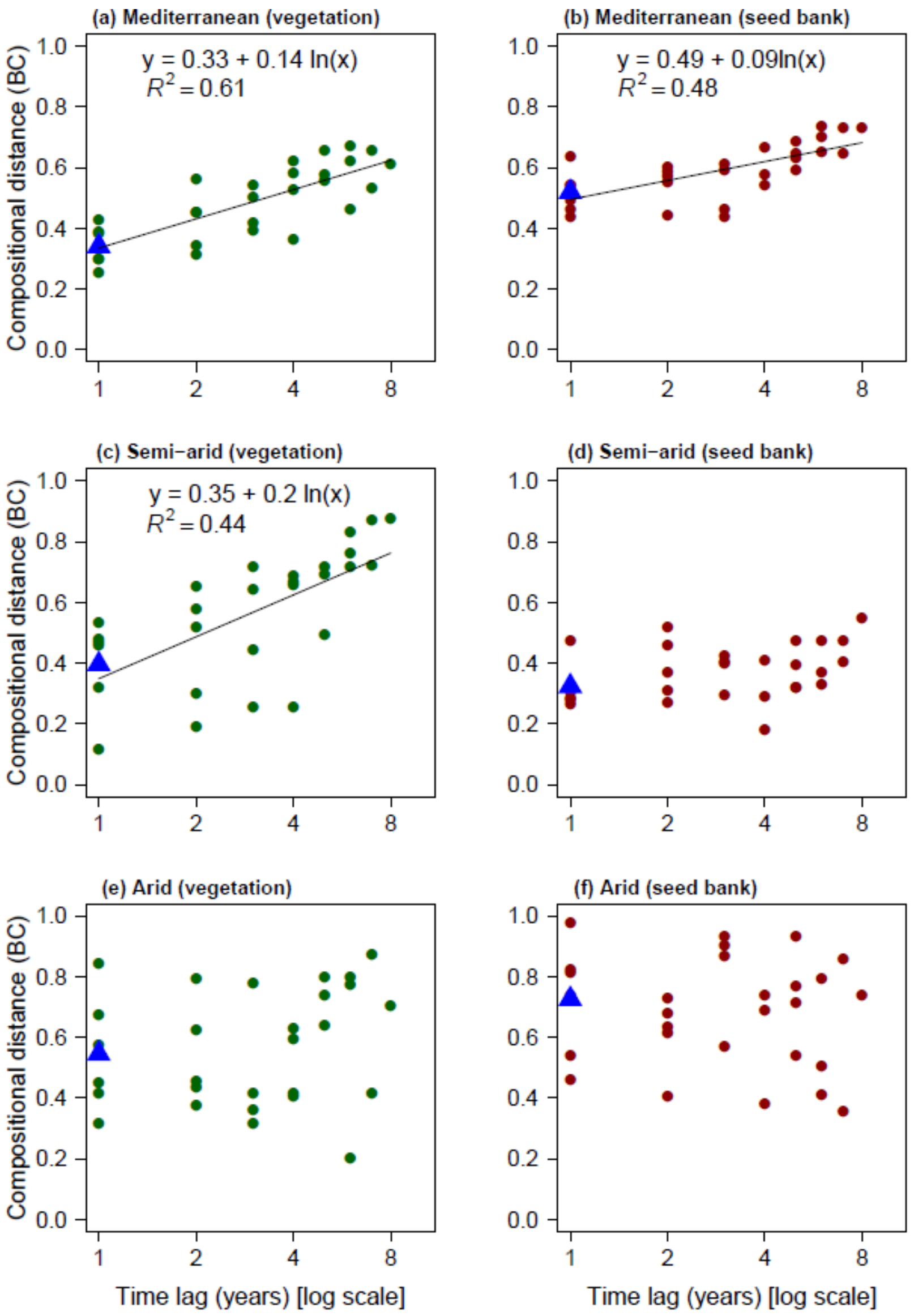

**Fig. S10:** Compositional distance (Bray–Curtis index) in the vegetation (left panels) and seed bank from *the first germination season* (right panels) as a function of time lag (temporal distance between years of sampling including all possible pairs). The blue triangle represents the mean compositional distance between two consecutive years (year-to-year variability). The slope of the relationship indicates the rate of long-term trends. (a,b) Mediterranean site, $P_{(\text{year-to-year variability})}<0.001$, $P_{(\text{slope})}<0.05$. (c,d) Semi-arid site, $P_{(\text{year-to-year variability})}>0.05$, $P_{(\text{slope})}< 0.001$. (e,f) Arid site, $P_{(\text{year-to-year variability})}>0.05$, $P_{(\text{slope})}>$ 0.05. The x-axis is on a logarithmic scale.

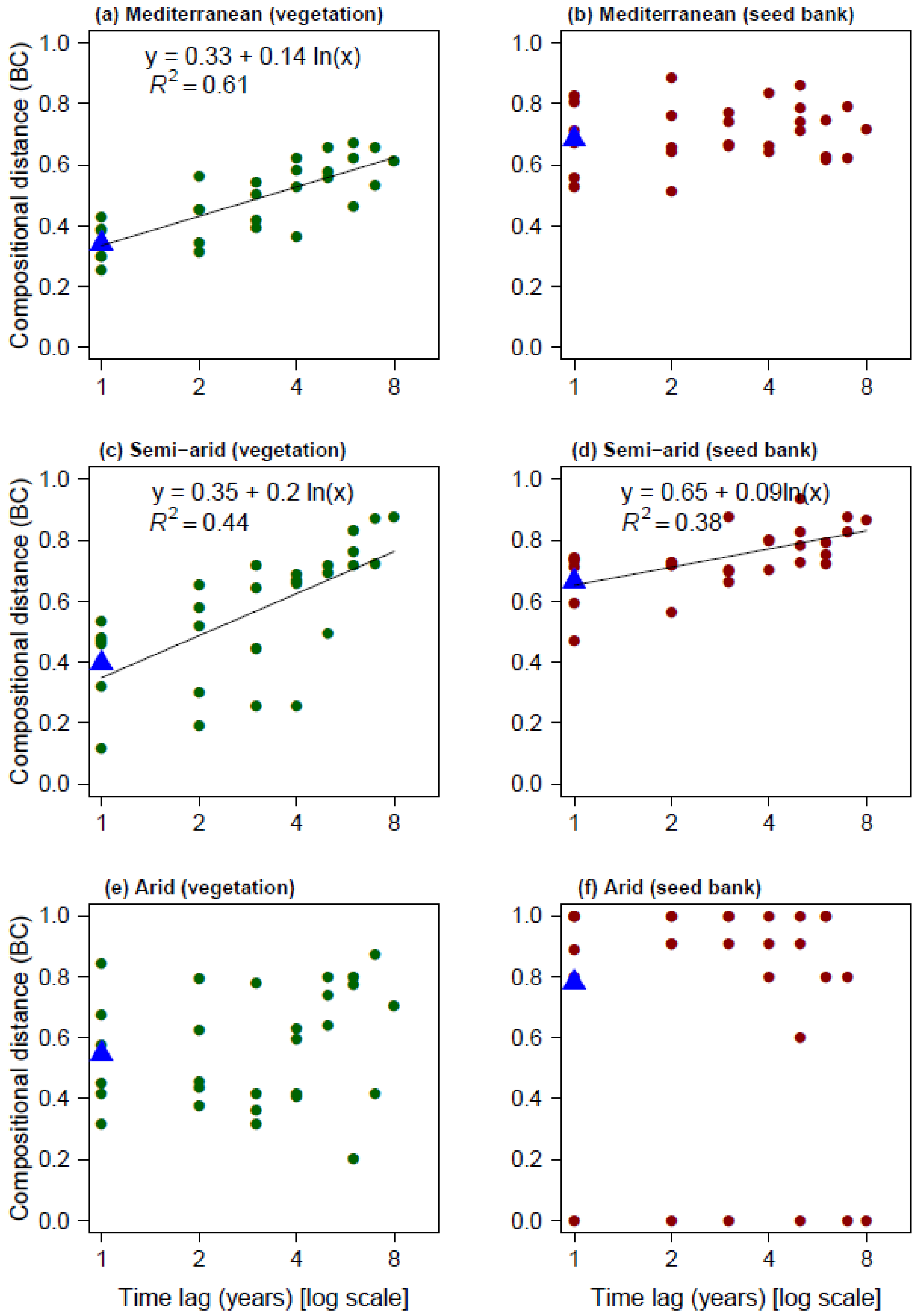

**Fig. S11:** Compositional distance (Bray–Curtis index) in the vegetation (left panels) and seed bank from *the second germination season* (right panels) as a function of time lag (temporal distance between years of sampling including all possible pairs). The blue triangle represents the mean compositional distance between two consecutive years (year-to-year variability). The slope of the relationship indicates the rate of long-term trends. (a,b) Mediterranean site, P(year-to-year variability)<0.05, P(slope)=0.01. (c,d) Semi-arid site, P(year-to-year variability)>0.05, P(slope)< 0.001. (e,f) Arid site, P(year-to-year variability)>0.05, P(slope)> 0.05. The x-axis is on a logarithmic scale.

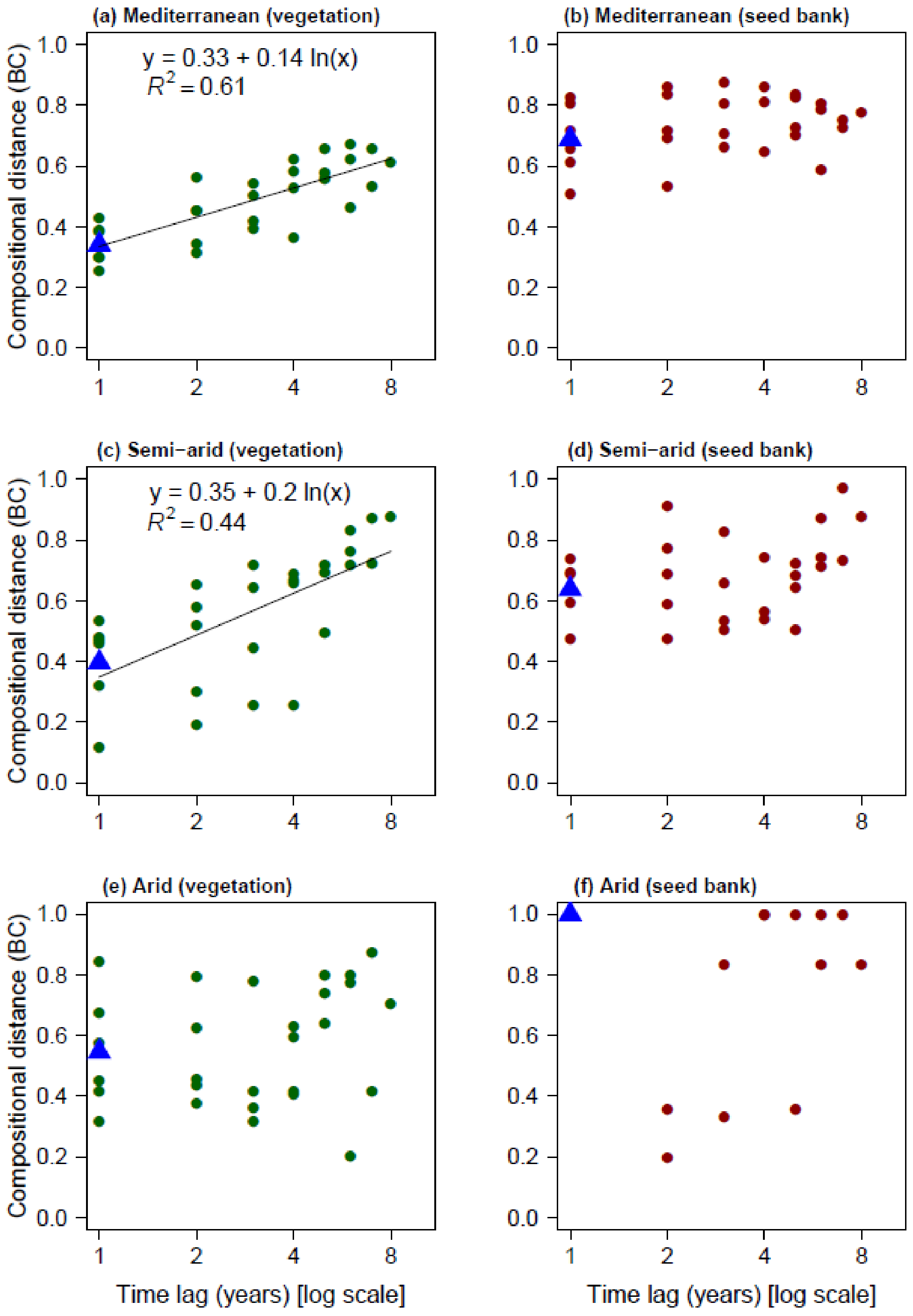

**Fig. S12:** Compositional distance (Bray–Curtis index) in the vegetation (left panels) and seed bank from *the third germination season* (right panels) as a function of time lag (temporal distance between years of sampling including all possible pairs). The blue triangle represents the mean compositional distance between two consecutive years (year-to-year variability). The slope of the relationship indicates the rate of long-term trends. (a,b) Mediterranean site, $P_{(\text{year-to-year variability})}<0.005$, $P_{(\text{slope})}<0.001$. (c,d) Semi-arid site, $P_{(\text{year-to-year variability})}<0.05$, $P_{(\text{slope})}< 0.005$ (e,f) Arid site, $P_{(\text{year-to-year variability})}<0.05$, $P_{(\text{slope})}>$ 0.05. The x-axis is on a logarithmic scale.

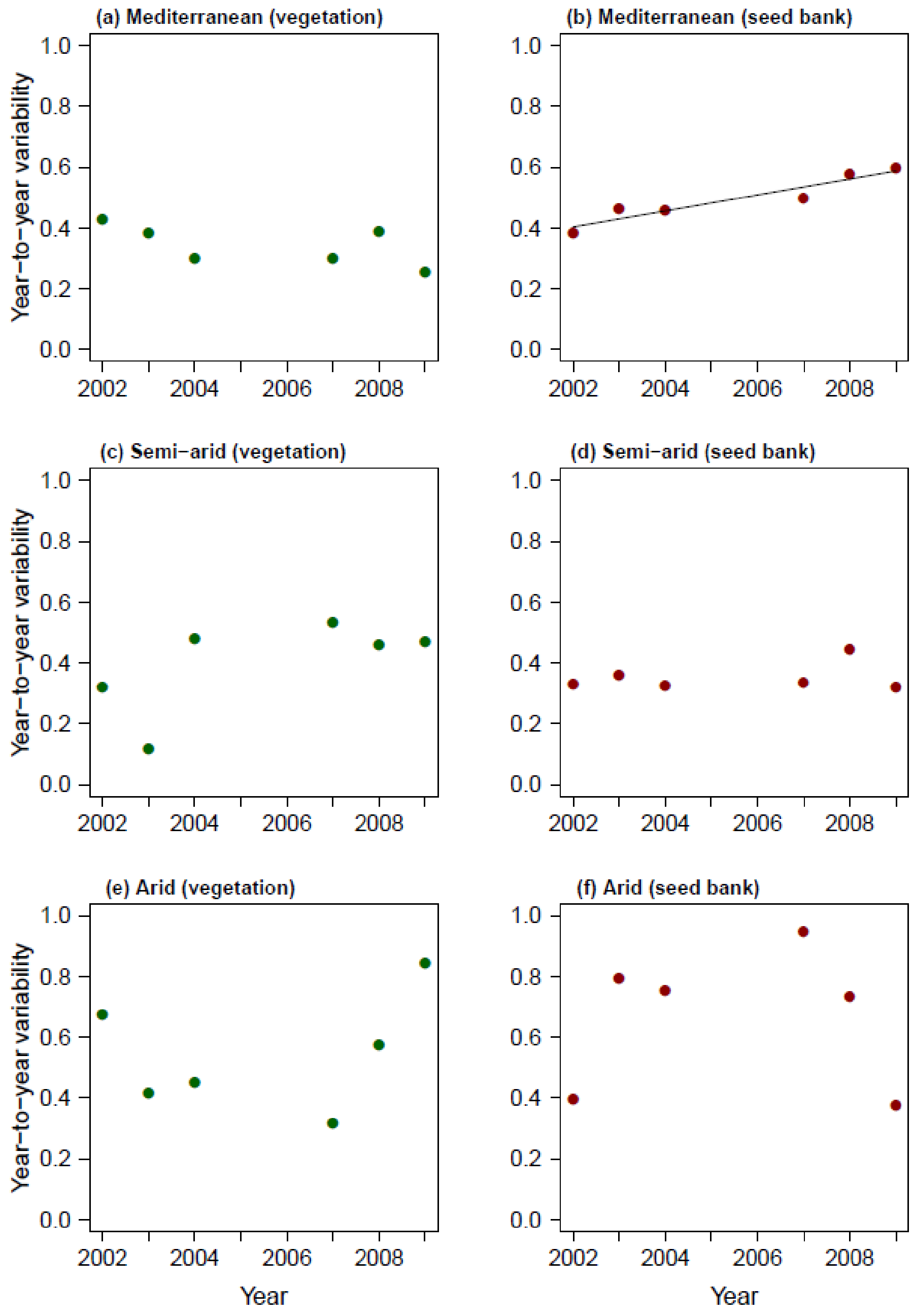

**Fig. S13.** Relationships between the year of sampling and year-to-year variability. Year-to-year variability is the Bray–Curtis distance between each year compared with the previous year (e.g. '2002' represents the distance between 2002\2003 and 2001\2002 growing seasons). A trend line appears when there is a significant linear trend ($P<0.05$).